\pdfoutput=1

\documentclass[twocolumn]{emulateapj}

\usepackage{amsmath}
\usepackage{graphicx}
\usepackage{color}
\usepackage[colorlinks]{hyperref}
\hypersetup{
    colorlinks,	
    citecolor=blue,
}

\newcommand{\be}{\begin{eqnarray}}
\newcommand{\ee}{\end{eqnarray}}

\shorttitle{Testing the Kerr black hole hypothesis with GRS~1915+105}
\shortauthors{Tripathi et al.}

\begin{document}

\title{Testing the Kerr black hole hypothesis with the continuum-fitting and the iron line methods: the case of GRS~1915+105}

\author{Ashutosh~Tripathi\altaffilmark{1}, Askar~B.~Abdikamalov\altaffilmark{1,2,3}, Dimitry~Ayzenberg\altaffilmark{4}, Cosimo~Bambi\altaffilmark{1,\dag}, Victoria~Grinberg\altaffilmark{5}, Honghui~Liu\altaffilmark{1}, and Menglei~Zhou\altaffilmark{6}}

\altaffiltext{1}{Center for Field Theory and Particle Physics and Department of Physics, 
Fudan University, 200438 Shanghai, China. \email[\dag E-mail: ]{bambi@fudan.edu.cn}}
\altaffiltext{2}{Ulugh Beg Astronomical Institute, Tashkent 100052, Uzbekistan}
\altaffiltext{3}{Tashkent Institute of Irrigation and Agricultural Mechanization Engineers, Tashkent 100000, Uzbekistan}
\altaffiltext{4}{Theoretical Astrophysics, Eberhard-Karls Universit\"at T\"ubingen, D-72076 T\"ubingen, Germany}
\altaffiltext{5}{European Space Agency (ESA), European Space Research and Technology Centre (ESTEC), Keplerlaan 1, 2201 AZ Noordwijk, The Netherlands}
\altaffiltext{6}{Institut f\"ur Astronomie und Astrophysik, Eberhard-Karls Universit\"at T\"ubingen, D-72076 T\"ubingen, Germany}

\begin{abstract}
The continuum-fitting and the iron line methods are currently the two leading techniques for probing the strong gravity region around accreting black holes. In the present work, we test the Kerr black hole hypothesis with the stellar-mass black hole in GRS~1915+105 by analyzing five disk-dominated \textsl{RXTE} spectra and one reflection-dominated \textsl{Suzaku} spectrum. The combination of the constraints from the continuum-fitting and the iron line methods has the potential to provide more stringent tests of the Kerr metric. Our constraint on the Johannsen deformation parameter $\alpha_{13}$ is $-0.15 < \alpha_{13} < 0.14$ at 3$\sigma$, where the Kerr metric is recovered when $\alpha_{13} = 0$.
\end{abstract}


\section{Introduction}

The \textit{Kerr hypothesis} states that all isolated, stationary, and axisymmetric astrophysical (uncharged) black holes (BHs) are described by the Kerr metric. This hypothesis is a consequence of the BH uniqueness theorems \citep[see, e.g.,][]{2012LRR....15....7C}. The Kerr metric is completely determined by two physical parameters: the mass of the BH $M$ and the spin angular momentum of the BH $|\vec{J}|$. The Kerr hypothesis holds in general relativity (GR) and some modified gravity theories~\citep{2008PhRvL.100i1101P}. However, there are theories in which the Kerr metric is not a solution \citep[see, e.g.,][]{Alexander:2009tp, Kleihaus:2011tg}. Additionally, the introduction of new physics, such as exotic matter fields~\citep{Herdeiro:2014goa} or macroscopic quantum effects~\citep{Giddings:2017jts}, can also lead to violations of the Kerr hypothesis. Tests of the Kerr hypothesis, and in turn GR, are necessary to gain a full understanding of BHs and gravity \citep{2011MPLA...26.2453B,2017RvMP...89b5001B}.

In this work, we test the Kerr hypothesis using a combined analysis with two BH electromagnetic observation methods: the continuum-fitting method \citep{1997ApJ...482L.155Z,2011CQGra..28k4009M,2014SSRv..183..295M} and X-ray reflection spectroscopy \citep{2006ApJ...652.1028B,2014SSRv..183..277R,2020arXiv201104792B}. The source of the electromagnetic radiation in the BH system is an accretion disk that surrounds the BH and, in the case studied here, is built up from gas that is siphoned off of a stellar companion. The accretion disk is generally assumed to be geometrically thin, optically thick, and cold (on the order of 1 keV for the stellar-mass BHs). The gas in the disk emits as a blackbody, but has a temperature that is dependent on its distance from the BH, and so the overall disk emits as a multi-temperature blackbody. This multi-temperature blackbody radiation is known as the BH continuum or thermal spectrum. These BH-disk systems also include a corona, which is a region of hotter ($\sim$100 keV) plasma that is close to the BH and the inner part of the accretion disk. Some of the thermal radiation from the disk inverse-Compton scatters off free electrons in the corona, producing a power-law spectrum with a high-energy cutoff. Some of this power-law emission illuminates the disk and is reflected, modifying the original power-law spectrum, most importantly with fluorescent emission lines in the soft X-ray band and a Compton hump that peaks at 20-30 keV~\citep{Ross:2005dm, Garcia:2010iz}. The brightest emission line is the iron K$\alpha$ complex at $\sim$6 keV and is known as the iron line. The overall reflected emission is known as the reflection spectrum and is studied through X-ray reflection spectroscopy.

The continuum spectrum and the reflection spectrum are both strongly dependent on the properties of the BH spacetime. Thus, these observations can in principle be used to test the Kerr hypothesis \citep{2002NuPhB.626..377T,2003IJMPD..12...63L,2009GReGr..41.1795S,2011ApJ...731..121B,2013ApJ...773...57J,2013PhRvD..87b3007B}. For this purpose, some of us have developed two {\tt XSPEC} models~\citep{1996ASPC..101...17A} to analyze observed spectra while allowing for a non-Kerr spacetime. The continuum spectrum is modeled through {\tt nkbb}~\citep{Zhou:2019fcg} and the reflection spectrum through {\tt relxill\_nk}~\citep{2017ApJ...842...76B, Abdikamalov:2019yrr}, which is an extension of the {\tt relxill} package~\citep{Dauser:2013xv, Garcia:2013lxa}. With these models, we can employ any well-behaved stationary and axisymmetric BH solution. However, here, and in past work, we primarily focus on solutions that parametrically deform the Kerr spacetime and do not represent any specific non-Kerr solution in GR or a modified gravity theory. By doing so we are performing a null test,~i.e.~determining if observations are consistent with a null result, which in our case is the Kerr solution and a test of the Kerr hypothesis \citep[see, e.g.,][]{2019ApJ...875...56T,2021ApJ...907...31T,2021ApJ...913...79T}. In this work, we use the Johannsen spacetime~\citep{Johannsen:2015pca}, which parametrically deforms the Kerr spacetime while still retaining three constants of the motion as in the Kerr solution. For simplicity, we here only study the strongest of the deformation parameters present in the Johannsen spacetime; however, there are an infinite number of deformation parameters present.

In this paper, we apply the continuum-fitting and iron line methods to the BH in GRS~1915+105. This source was discovered by \textsl{Granat} satellite in 1992 \citep{1992IAUC.5590....2C}. It is identified as a microquasar, as it displays radio jets \citep{1999ARA&A..37..409M}. GRS~1915+105 shows remarkable variability across the whole electromagnetic spectrum \citep{2000A&A...355..271B, 2002MNRAS.331..745K}. \citet{1997ApJ...482L.155Z} reported a very high spin of this source and claimed it to be the reason for its exceptional variability and presence of relativistic jets. Observation from \textsl{BeppoSAX} clearly identified the broad Iron K$\alpha$ line around 6.4~keV \citep{2002A&A...387..215M}. The most recent measurements of the mass and distance of this source are $12.4^{+2.0}_{-1.8}$ $M_{\odot}$ and $8.6^{+2.0}_{-1.6}$~kpc, respectively \citep{2014ApJ...796....2R}. The broad line emission has also been observed with other X-ray missions and high spin measurements are reported using the continuum fitting method for \textsl{RXTE} data \citep{2006ApJ...652..518M} and X-ray reflection spectroscopy for \textsl{Suzaku} \citep{2009ApJ...706...60B} and \textsl{NuSTAR} observations \citep{2013ApJ...775L..45M}. \citet{2019ApJ...875...41Z,2019ApJ...884..147Z} analyzed the \textsl{NuSTAR} and \textsl{Suzaku} observations, respectively, to test GR by employing the Johannsen metric.

This paper is organized as follows. In Section~\ref{sec:models}, we briefly describe the {\tt relxill\_nk} and {\tt nkbb} models. In Section~\ref{sec:analysis}, we describe the \textsl{RXTE} and \textsl{Suzaku} observations, our data reduction, and the spectral analysis with {\tt relxill\_nk} and {\tt nkbb}. We discuss our results and conclude in Section~\ref{sec:discussion}. The Johannsen metric employed in our spectral analysis is reported in the appendix.


\section{The {\tt relxill\_nk} and {\tt nkbb} models}
\label{sec:models}

Here we briefly review the {\tt relxill\_nk} and {\tt nkbb} models. Since ray-tracing calculations, which give us the spectrum of a disk away from the source, require too much time and resources for the real-time data analysis stage, both models use the transfer function formalism proposed by Cunningham~\citep{1975ApJ...202..788C}, which allows us to tabulate all the necessary details of the spacetime metric to a FITS (Flexible Image Transfer System) file. The tabulated data can then be extracted for use in the data analysis phase.

With the transfer function formalism, the observed flux can be written as 
\be
    F_{o}(E_{o})&=&\frac{1}{D^2}\int^{R_{\text{out}}}_{R_{\text{in}}}
    \int^{1}_{0}\frac{\pi r_{e}g^{2}f(g^{*},r_{e},\iota)}{\sqrt{g^{*}(1-g^{*})}}
    \nonumber\\
    && \qquad \qquad \qquad \times I_{e}(E_e, r_{e},\theta_{e})dg^{*}dr_{e}, \label{eq:flux}
\ee
where $D$ denotes the distance from the source to the observer's screen, $R_{\rm in}$ ($R_{\rm out}$) is the inner (outer) boundary of the accretion disk, $r_e$ is the radial coordinate of the emitting rings in the accretion disk, $g = E_o/E_e$ is the redshift factor, defined as the ratio of the energy of photons registered on the observer's screen ($E_o$) and at the emission point in the rest frame of the disk's gas ($E_e$). $I_e(E_e, r_e, \theta_e)$ is the specific intensity at the emission point on the accretion disk at the energy $E_e$, the radial coordinate $r_e$, and the radiation angle $\theta_e$ (angle between the emitted photons and the normal to the disk). The transfer function $f$ has the following form
\begin{equation}
f(g^{*},r_{e},\iota)=\frac{1}{\pi r_{e}}g\sqrt{g^{*}(1-g^{*})}\left|\frac{\partial(X,Y)}{\partial(g^{*},r_{e})}\right|.
\end{equation}
Here $\left|\partial(X,Y)/\partial(g^{*}, r_{e})\right|$ represents the Jacobian, $X$ and $Y$ are the Cartesian coordinates in the observer's screen, and $g^ {*}$ is the relative redshift factor, which is written as
\begin{equation}
g^{*}=\frac{g-g_{\text{min}}}{g_{\text{max}}-g_{\text{min}}}\in[0,1].
\end{equation}
where $g_{\rm min}=g_{\rm min}(r_e, \iota)$ ($g_{\rm max}=g_{\rm max}(r_e, \iota)$) denotes the minimum (maximum) value of the redshift factor $g$ for a given radial coordinate $r_e$ and the viewing angle $\iota$ of an observer.
Thus, the transfer function formalism can separate the calculations related to the spectrum appearing on the surface of the disk in the rest-frame of the emitting gas from those that transmit the emitted spectrum from the emission point to a distant observer, taking into account the effects of the spacetime metric. This separation results in a fast calculation of the observed spectrum without the need to recalculate all the photon trajectories.

In both models, the transfer functions are pre-calculated for a three-dimensional grid, which consists of the BH spin parameter $a_*$, the Johannsen spacetime with deformation parameter $\alpha_{13}$~\citep{Johannsen:2015pca}, and the viewing angle $\iota$, and are tabulated in the FITS table. The parameter ranges of the grid are selected so that the entire parameter space is effectively covered. At the data analysis stage, the transfer function values corresponding to the set of model input parameters are passed to Eq.~\ref{eq:flux} using interpolation methods.

{\tt relxill\_nk} adopts the model suite {\tt xillver}~\citep{Garcia:2010iz, Garcia:2013lxa} for the calculation of the reflection spectrum of the accretion disk at the emission point, namely for calculating $I_{e}(E_e, r_{e},\theta_{e})$. Therefore, all model parameters of {\tt xillver} are included in {\tt relxill\_nk}.

{\tt nkbb} assumes that the accretion disk is in local thermal equilibrium, and the emission at each point of the disk is like a blackbody. The local specific intensity of the disk radiation is taken as (in units in which $h = c = k_{\rm B} = 1$)
\begin{equation}
I_e(E_e)=\frac{2 E^3_e}{f^4_{\rm col}}
\frac{\Upsilon}{\exp{\left(\frac{E_e}{T_{\rm col}}\right)} - 1},
\end{equation}
where $f_{\rm col}$ is the color factor (or hardening factor), which is introduced to account for non-thermal effects occurring mainly due to electron scattering in the disk atmosphere. $\Upsilon$ is a function of the angle between the emitted photons from the disk surface and the normal to the disk. Isotropic emission ($\Upsilon=1$) and limb-darkened emission ($\Upsilon=\frac{1}{2}+\frac{3}{4}\cos\theta_e$) are the two most common options. $T_{\rm col}(r)=f_{\rm col} T_{\rm eff}$ is the color temperature, and $T_{\rm eff}(r)$ is the effective temperature at every radius defined from the relation $\mathcal{F}(r)=\sigma T^4_{\rm eff}$, where $\sigma$ denotes the Stefan-Boltzmann constant and $\mathcal{F}(r)$ is the time-averaged energy flux emitted by the surface of the Novikov-Thorne disk defined as \citep[see, e.g.,][]{2012ApJ...761..174B}
\begin{equation}
\mathcal{F}(r) = \frac{\dot{M}}{4\pi M^2}F(r).
\end{equation}
Here $M$ and $\dot{M}$ are, respectively, the BH mass and the mass accretion rate, and $F(r)$ is a dimensionless function that depends on the spacetime metric. In the end, the BH mass $M$, the mass accretion rate $\dot{M}$, the BH distance $D$, the BH spin $a_*$, the Johannsen spacetime deformation parameter $\alpha_{13}$, the observer's viewing angle $\iota$, the function $\Upsilon$, and the color factor $f_{\rm col}$ are the parameters of the {\tt nkbb} model.

\begin{figure*}
\begin{center}
\includegraphics[width=8.5cm,trim={1.0cm 0cm 4.0cm 16cm},clip]{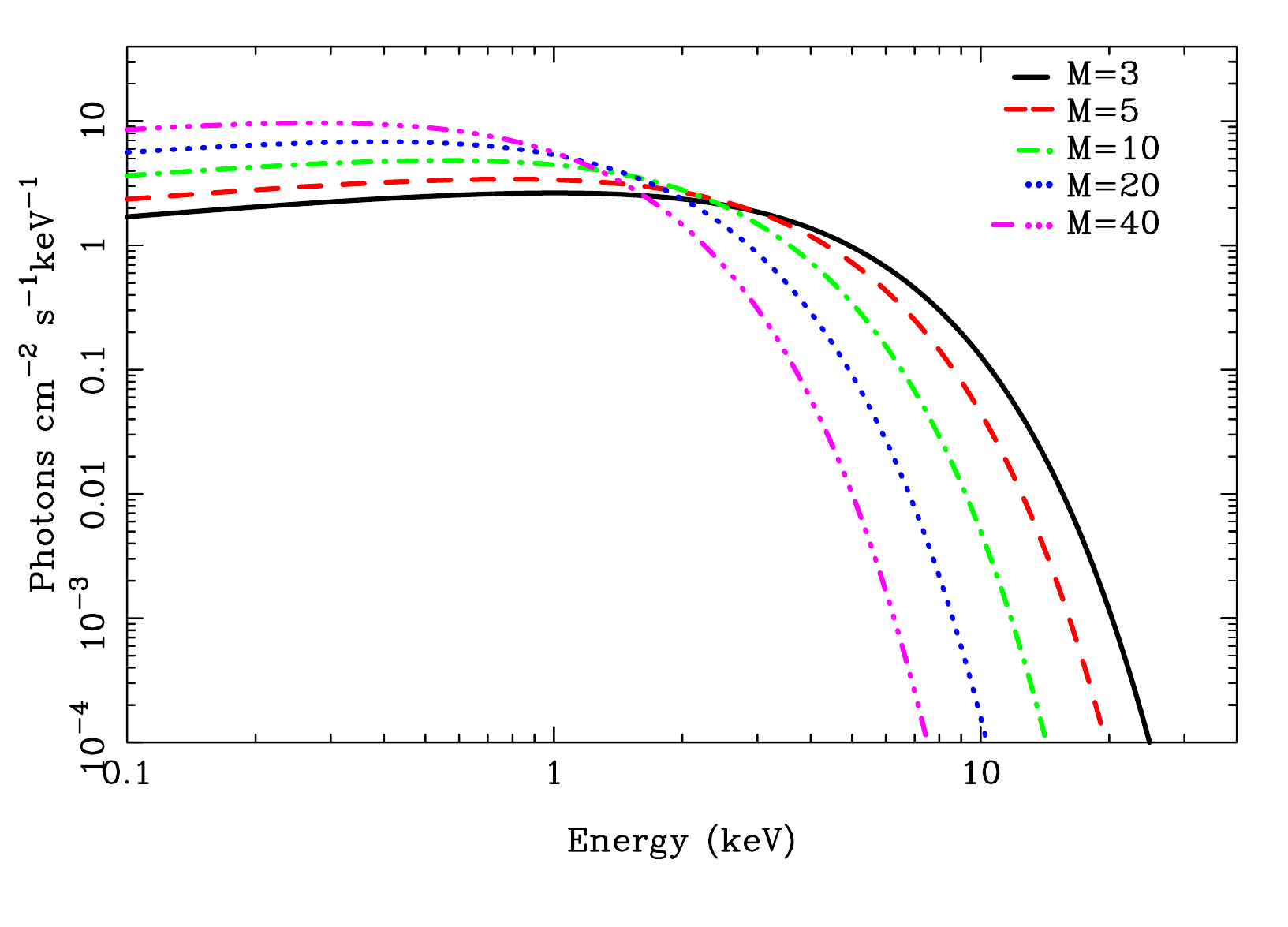}
\hspace{0.5cm}
\includegraphics[width=8.5cm,trim={1.0cm 0cm 4.0cm 16cm},clip]{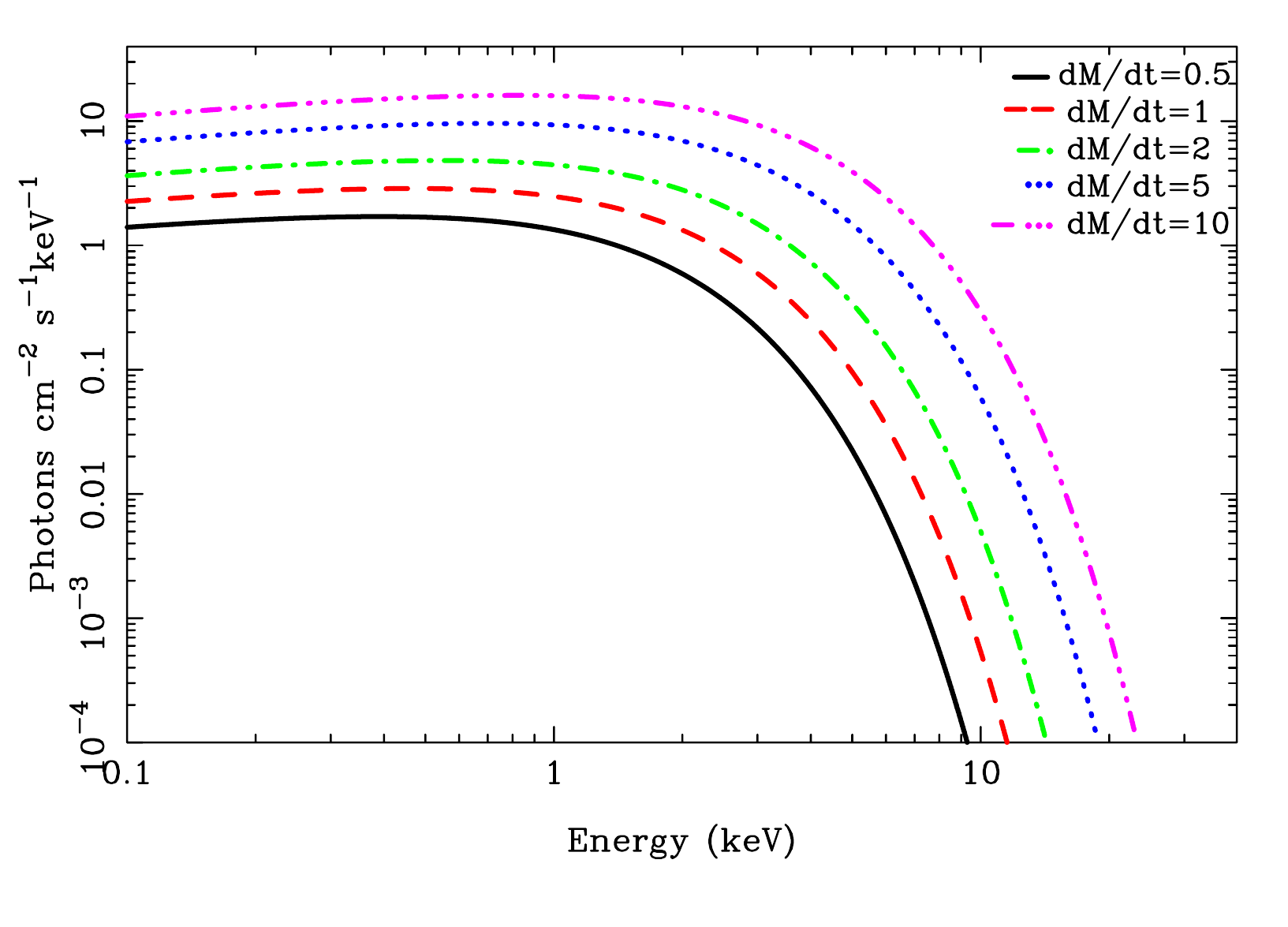} \\
\includegraphics[width=8.5cm,trim={1.0cm 0cm 4.0cm 16cm},clip]{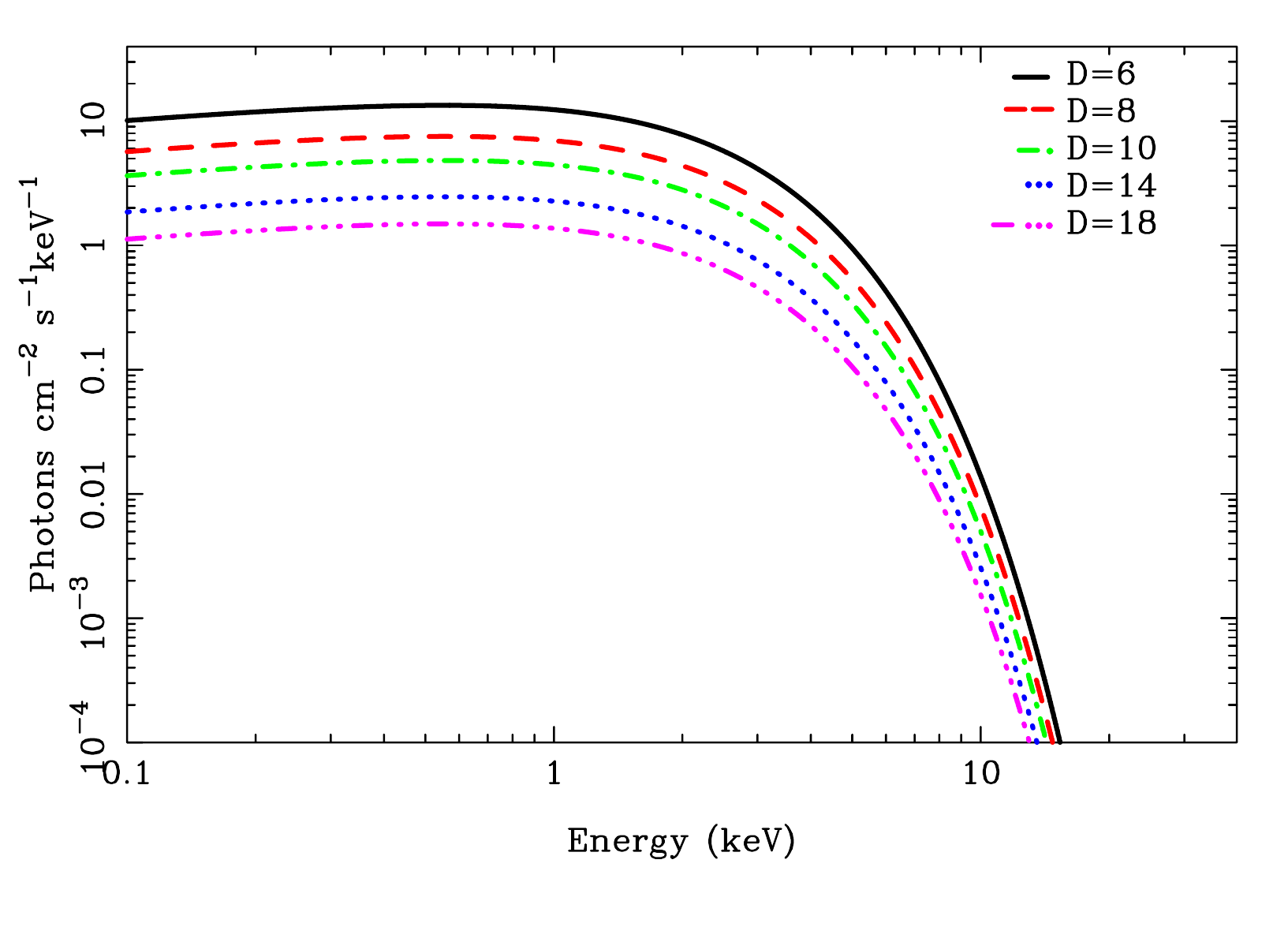}
\hspace{0.5cm}
\includegraphics[width=8.5cm,trim={1.0cm 0cm 4.0cm 16cm},clip]{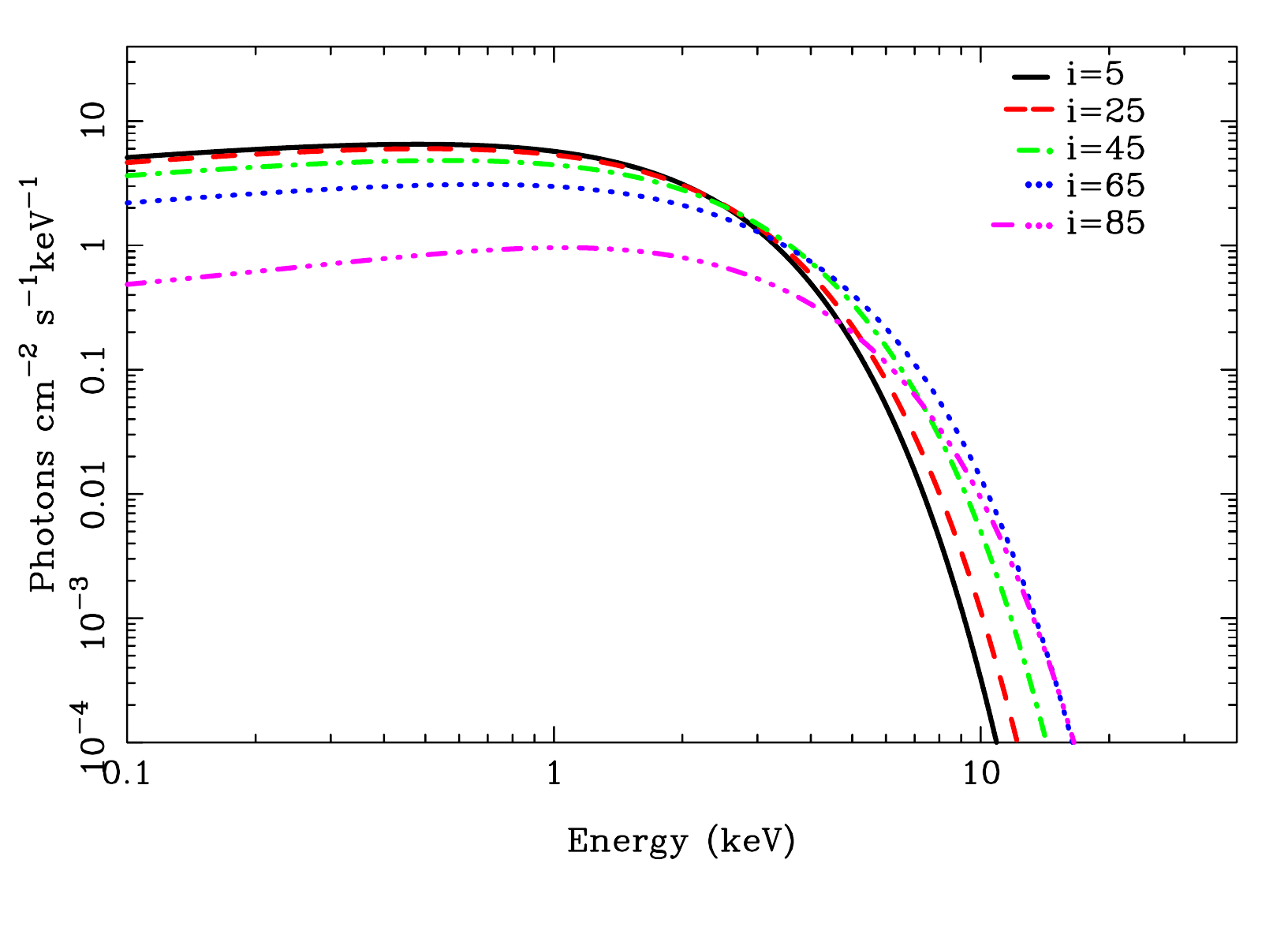} \\
\includegraphics[width=8.5cm,trim={1.0cm 0cm 4.0cm 16cm},clip]{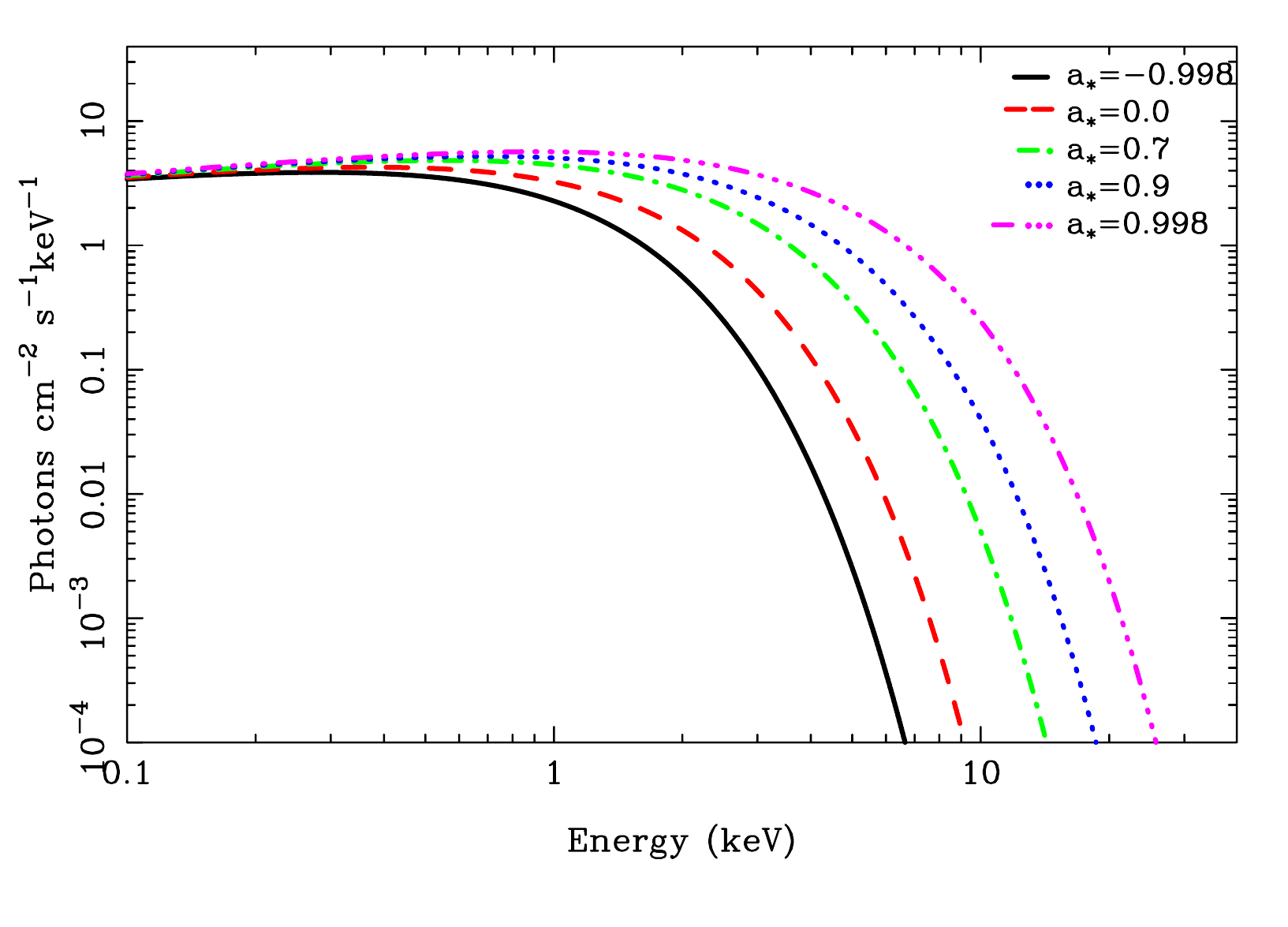}
\hspace{0.5cm}
\includegraphics[width=8.5cm,trim={1.0cm 0cm 4.0cm 16cm},clip]{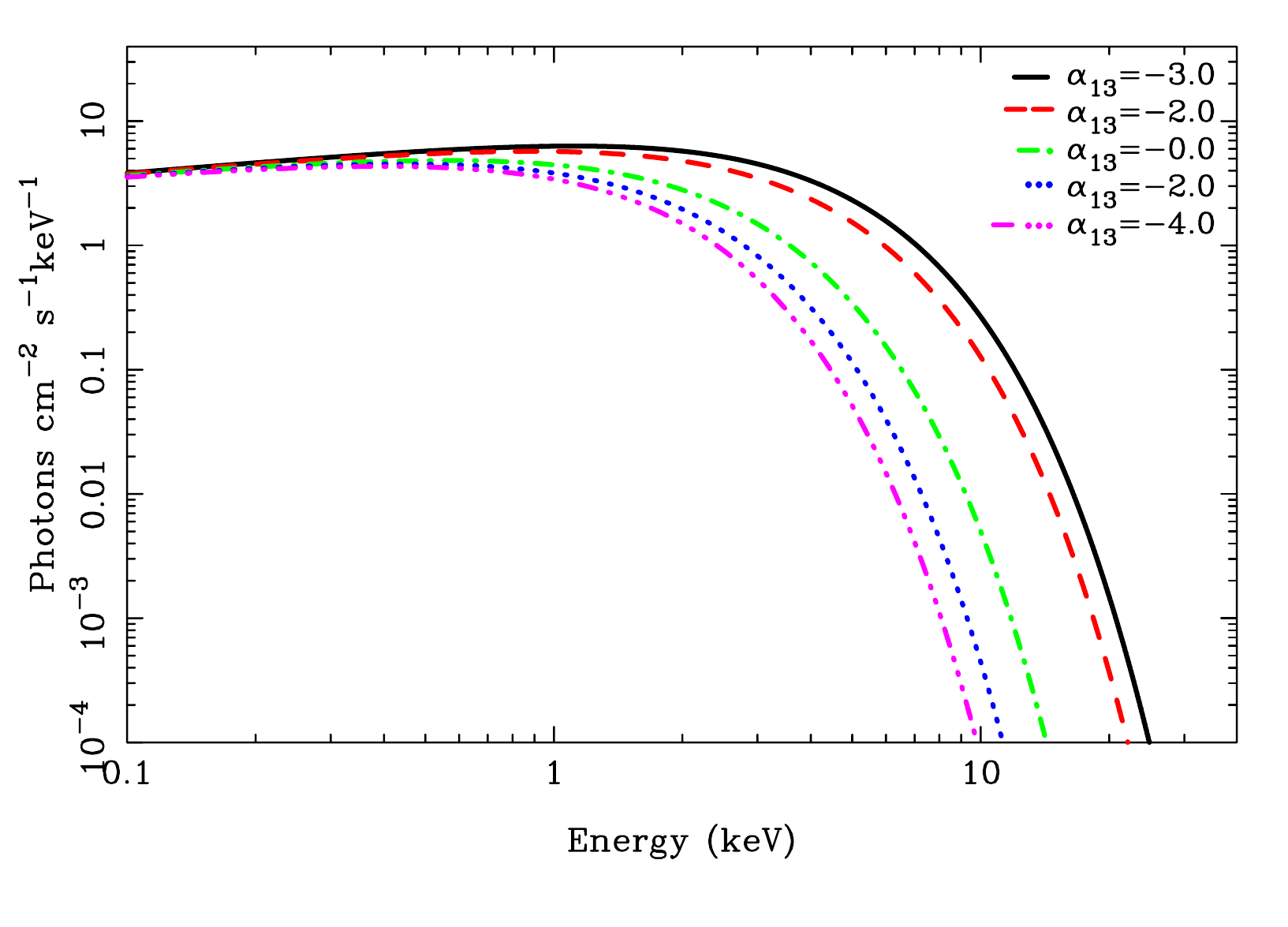}
\end{center}
\vspace{-0.5cm}
\caption{{Impact of the BH mass $M$, the mass accretion rate $\dot{M}$, the distance $D$, the inclination angle of the disk $\iota$, the BH spin parameter $a_*$, and the deformation parameter $\alpha_{13}$ on the thermal spectra of accretion disks as calculated by {\tt nkbb}. $M$ is in units of $M_\odot$; $\dot{M}$ is in units of $10^{18}$~g/s; $D$ is in units of kpc; $\iota$ in deg. For all spectra, we assume $f_{\rm col} = 1.7$ and $\Upsilon = 1$. The default values of the model parameters are: $M = 10$~$M_\odot$, $\dot{M}=2 \cdot 10^{18}$~g/s, $D = 10$~kpc, $\iota = 45^\circ$, $a_* = 0.7$, and $\alpha_{13} = 0$.} 
\label{f-nkbb}}
\end{figure*}

Fig.~\ref{f-nkbb} shows how the parameters $M$, $\dot{M}$, $D$, $\iota$, $a_*$, and $\alpha_{13}$ affect the thermal spectrum. It is clear that the BH spin parameter $a_*$ and the Johannsen deformation parameter $\alpha_{13}$ have a very similar impact on the shape of the spectrum. The high-energy cutoff of the spectrum is indeed determined by the inner edge of the disk, which is assumed to be at the innermost stable circular orbit (ISCO). When the BH mass is known, the ISCO is determined by $a_*$ and $\alpha_{13}$. Apart from determining the ISCO radius, the impact of $a_*$ and $\alpha_{13}$ on the thermal spectrum of the disk is very weak. As already pointed out in \citet{2014ApJ...797...78K}, in such a situation there is a strong correlation between the estimates of the spin and the deformation parameters and -- as we will verify in the next section -- our constraints on the plane $a_*$ vs. $\alpha_{13}$ will appear with the typical banana-shape of two degenerate parameters.

\section{Data reduction and analysis}
\label{sec:analysis}

{While GRS~1915+105 has been observed many times and by different X-ray missions, there are not many observations of this source suitable for the continuum-fitting method and the analysis of the reflection features of the disk. In part, this is because GRS~1915+105 is quite a peculiar and complicated source. Its mass accretion rate is often too high for maintaining a thin accretion disk, so our models based on Novikov-Thorne disks are not applicable. To the best of our knowledge, there are only the \textsl{RXTE} data studied in \citet{2006ApJ...652..518M} suitable for the continuum-fitting method and the \textsl{Suzaku} observation analyzed in \citet{2009ApJ...706...60B} suitable for X-ray reflection spectroscopy. These will be the data analyzed in our study.}

\subsection{\textsl{RXTE} data}

For data selection, we follow \citet{2006ApJ...652..518M} which used data taken with the Proportional Counter Array (PCA) onboard the Rossi X-ray Timing Explorer (\textsl{RXTE}) mission in which the source is in a thermal dominant state. A thermal dominant state \citep{2006ARA&A..44...49R} is defined as a period with a majority ($>75\%$) of the emission from the accretion disk over the 2.0-20.0 keV energy band, low root mean square (RMS) variability ($<0.075$ integrated over 0.1-1.0 Hz) and absence or very weak presence of quasi-periodic oscillations (QPOs) (amplitude less than $0.005\%$). By screening all \textsl{RXTE} archival data of this source with thermal dominance criteria, \citet{2006ApJ...652..518M} identified 20 observations for further analysis. Out of these 20 observations, we select five key low-luminosity observations. This is because, for the continuum-fitting method used in this work, the luminosity should be less than 30\% of the Eddington luminosity. Tab.~\ref{t-obs} shows the basic details of the five \textsl{RXTE} observations used in this work.

\subsubsection{Data reduction}

For data reduction, we follow \citet{2020MNRAS.496..497Z} which uses a standard data reduction approach. We only include data from PCU-2 because it is the best calibrated PCU and was always operating. We extract the PCA spectra in the {\tt standard2f} mode. We also ignore the data taken within 10~minutes of the South Atlantic Anomaly (SAA). Using {\tt pcacorr} tool \citep{2014ApJ...794...73G}, the calibration of the PCA data is further improved. As the count rates are high, the systematic uncertainty of 0.1\% is added to all PCA data following \citet{2014ApJ...794...73G}. We rebin the data such that the signal-to-noise (S/N) ratio is 10. We use data in the energy range 3.0-45.0 keV for all five observations for further spectral analysis.

\subsubsection{Data analysis}

The spectral analysis work in this paper is done using {\tt XSPEC} v12.11.1 \citep{1996ASPC..101...17A}. 
{We consider the abundances from \citet{2000ApJ...542..914W} and the cross-sections from \citet{1996ApJ...465..487V}.} For galactic absorption, {\tt tbabs} \citep{2000ApJ...542..914W} is used with column density frozen at $8 \times 10^{22}$ cm$^{-2}$ following \citet{2020ApJ...899...80A}. We note that the exact value of the column density has a very weak impact on the final fit. We start our analysis by fitting the data with an absorbed power-law continuum ({\tt powerlaw}) and a multi-temperature disk blackbody ({\tt diskbb}) \citep{1984PASJ...36..741M}. There are several narrow absorption features present in the energy range of 6.4-8.4 keV which is also consistent with \citet{2000ApJ...539..413K}. Recently, such narrow absorption lines have also been observed with \textsl{NICER}: \citet{2020ApJ...902..152N} reported the detection of multiple absorption features in the energy range 6.4-8.4~keV. {Such absorption features can be interpreted as different iron absorption lines. In our case, given the limited resolution of the PCA spectra, we follow \citet{2006ApJ...652..518M} and we add a broad Gaussian absorption line ({\tt gabs}) with a limited width of 0.5 keV and energy restricted between 6.3 and 7.5 keV.} 

We also add {\tt smedge} \citep{1994PASJ...46..375E} to model the broad Fe absorption edge with the energies to vary between 6.9 and 9.0 keV. The addition of these two components significantly improves the fit.

After identifying the line components, we replace the Newtonian disk model {\tt diskbb} with {\tt nkbb} \citep{Zhou:2019fcg}, which calculates relativistic thermal spectra of accretion disks in the Johannsen metric. The spacetime parameters are the BH mass, the (dimensionless) BH spin $a_*$, and the deformation parameter $\alpha_{13}$. The other model parameters are the distance ($D$), the inclination angle of the disk with respect to the line of sight of an observer ($\iota$), and the mass accretion rate ($\dot{M}$). Due to degeneracies, it is not possible to measure these parameters simultaneously by fitting the data of a source. So, we need independent measurements of $M$, $\iota$, and $D$. The fit will provide the measurements of $a_*$, $\dot{M}$, and $\alpha_{13}$. We freeze the mass of the source at $M = 12.4~M_{\odot}$ and distance at $D = 8.6$~kpc \citep{2014ApJ...796....2R}. Here, we freeze the value of inclination angle $\iota$ to 65$^{\circ}$ measured by high resolution radio images \citep{1999MNRAS.304..865F} assuming that the jet of the source is parallel to the BH spin and orthogonal to the accretion disk. We assume limb darkening emission ({\tt lflag=1}). We freeze the spectral hardening factor $f_{\rm col}$ to 1.7, which is the most widely used value for a stellar-mass BH with an Eddington-scaled accretion luminosity around 10\%  \citep{1995ApJ...445..780S}. As we freeze $M$, $\iota$, and $D$, the normalization of {\tt nkbb} is frozen to 1.

We found that, in some cases, the power-law continuum contributes significantly to the flux at energies below 5 keV. This should not be the case as the power-law emission is produced by Comptonization of soft disk photons in the corona. So, we consider more realistic models. We try {\tt nthcomp}, {\tt comptt}, and {\tt simplcutx}. We find that {\tt comptt} \citep{1994ApJ...434..570T} always fits the data better than the other two models, so we use it to describe the coronal spectrum for further analysis.

As pointed out in \citet{2006ApJ...652..518M}, these low luminosity observations give high $\chi^2$ values ($\sim$1.5). In order to improve the fit, we follow \citet{2006ApJ...652..518M} and add a sharp edge feature ({\tt edge} in {\tt XSPEC}) with energy allowed to vary between 8 and 13 keV and we refit the spectra. Note that {\tt edge} improves $\chi^2$ statistics, but it does not have any significant impact on the measurement of $a_*$, $\dot{M}$, and $\alpha_{13}$, in agreement with \citet{2006ApJ...652..518M}. In {\tt XSPEC} jargon, the model eventually reads as 

\vspace{0.2cm}

{\tt tbabs$\times$smedge$\times$gabs$\times$edge$\times$(comptt+nkbb)} .

\vspace{0.2cm}

\begin{figure}
\begin{center}
\includegraphics[width=8.5cm,trim={2.0cm 2.0cm 3.0cm 10.0cm},clip]{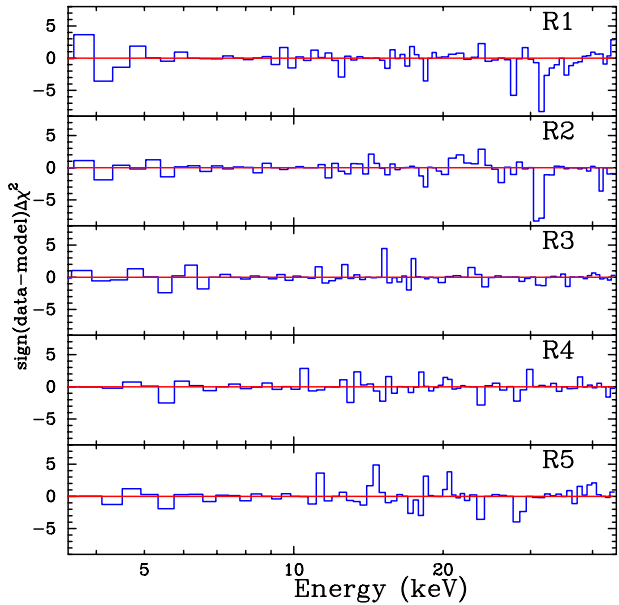}
\end{center}
\vspace{-0.2cm}
\caption{Residuals of the individual fits of the five \textsl{RXTE} observations of GRS~1915+105. \label{f-R-r}}
\end{figure}

The results of our fits are summarized in Tab.~\ref{t-rxte}, which shows both the measurements of the fits of the individual observations (R1, R2, R3, R4, and R5) and of the fit of all observations together (R1-R5, last column). Fig.~\ref{f-R-r} shows the residuals of the individual fits.

\begin{table*}
\centering
{\renewcommand{\arraystretch}{1.2}
\begin{tabular}{lccccc}
\hline\hline
Obs. & \hspace{0.3cm} Mission \hspace{0.3cm} & \hspace{0.3cm} Observation ID \hspace{0.3cm} & \hspace{0.3cm} Observation Date \hspace{0.3cm} & \hspace{0.3cm} Exposure (ks) \hspace{0.3cm} & \hspace{0.3cm} Counts/s \hspace{0.3cm} \\
\hline\hline
R1 & \textsl{RXTE} & 10408-01-20-00 & 1996 Jul 3 & 3.3 &  1349 \\
R2 & \textsl{RXTE} & 10408-01-20-01 & 1996 Jul 3 & 2.9 & 1381 \\
R3 & \textsl{RXTE} & 30703-01-13-00 & 1998 Mar 29 & 4.7 & 1230 \\
R4 & \textsl{RXTE} & 70702-01-36-00 & 2003 Jan 1 & 2.3 & 1518 \\
R5 & \textsl{RXTE} & 80701-01-34-00 & 2003 Nov 24 & 3.2 & 1496 \\   
\hline
S1 & \textsl{Suzaku} & 402071010 & 2007 May 27 & 28.9 & 84 \\
\hline\hline
\end{tabular}
}
\caption{\rm Summary of the six observations of GRS~1915+105 analyzed in the present study. \label{t-obs}}
\end{table*}

\begin{table*}
\centering
{\renewcommand{\arraystretch}{1.2}
\begin{tabular}{lcccccc}
\hline\hline
Parameter & \hspace{0.4cm} R1 \hspace{0.4cm} & \hspace{0.4cm} R2 \hspace{0.4cm} & \hspace{0.4cm} R3 \hspace{0.4cm} & \hspace{0.4cm} R4 \hspace{0.4cm} & \hspace{0.4cm} R5 \hspace{0.4cm} & \hspace{0.4cm} R1-R5 \hspace{0.4cm} \\
\hline
{\tt tbabs} & \\
$N_{\rm H}$ [$10^{22}$~cm$^{-2}$] & $8^\star$ & $8^\star$ & $8^\star$ & $8^\star$ & $8^\star$ & $8^\star$ \\
\hline
{\tt smedge} & \\
$E_{\rm edge}$ [keV] & $7.64_{-0.22}^{+0.22}$ & $7.63_{-0.18}^{+0.27}$ & $7.52_{-0.17}^{+0.14}$ & $7.7_{-0.3}^{+0.5}$ & $7.39_{-0.24}^{+0.20}$ \\
$\tau_{\rm max}$ & $0.73_{-0.32}^{+0.14}$ & $0.8_{-0.3}^{+0.3}$ & $1.07_{-0.21}^{+0.26}$ & $0.24_{-0.04}^{+0.22}$ & $0.9_{-0.3}^{+0.6}$ \\
W [keV] & $0.97_{\rm - (P)}^{+0.03}$ & $1.0_{\rm - (P)}^{+0.4}$ & $1.6_{-0.7}^{+0.8}$ & $0.2^{+0.2}$ & $2.0_{-0.8}^{+1.7}$ \\
\hline
{\tt gabs} & \\
$E_{\rm line}$ [keV] & $7.11_{-0.07}^{+0.06}$ & $7.13_{-0.08}^{+0.07}$ & $7.11_{-0.08}^{+0.07}$ & $7.14_{-0.12}^{+0.04}$ & $7.09_{-0.13}^{+0.11}$ \\
\hline
{\tt edge} & \\
$E_{\rm edge}$ [keV] & $9.0_{-0.4}^{+0.3}$ & $9.1_{-0.4}^{+0.4}$ & $8.84_{-0.18}^{+0.22}$ & $8.99_{-0.11}^{+0.04}$ & $8.90_{-0.22}^{+0.34}$ \\
$\tau_{\rm max}$ & $0.08_{-0.05}^{+0.06}$ & $0.08_{-0.05}^{+0.06}$ & $0.14_{-0.06}^{+0.05}$ & $0.189_{-0.007}^{+0.007}$ & $0.07_{-0.04}^{+0.04}$ \\
\hline
{\tt comptt} & \\
$kT_0$ [keV] & $0.32_{-0.32}^{+0.11}$ & $0.36_{-0.12}^{+0.05}$ & $0.54_{-0.05}^{+0.04}$ & $0.2_{\rm - (P)}^{+3.4}$ & $0.56_{-0.07}^{+0.03}$ \\
$kT_{\rm e}$ [keV] & $4.6_{-0.7}^{+1.6}$ & $4.7_{-1.1}^{+4.6}$ & $6.4_{-2.3}^{+29.2}$ & $3.0_{-0.4}^{+0.4}$ & $10_{-6}^{+106}$ \\
$\tau$ & $2.7_{-0.7}^{+0.6}$ & $2.6_{-0.5}^{+0.7}$ & $1.7_{-1.0}^{+1.8}$ & $4.72_{-0.02}^{+0.94}$ & $1.1_{-0.6}^{+1.8}$ \\
\hline
{\tt nkbb} & \\
$M$ [$M_\odot$] & $12.4^\star$ & $12.4^\star$ & $12.4^\star$ & $12.4^\star$ & $12.4^\star$ & $12.4^\star$ \\
$D$ [kpc] & $8.6^\star$ & $8.6^\star$ & $8.6^\star$ & $8.6^\star$ & $8.6^\star$ & $8.6^\star$ \\
$i$ [deg] & $65^\star$ & $65^\star$ & $65^\star$ & $65^\star$ & $65^\star$ & $65^\star$ \\
$a_*$ & $0.962_{-0.016}^{\rm +(P)}$ & $0.975_{-0.013}^{\rm +(P)}$ & $0.998_{-0.024}$ & $0.94_{-0.03}^{\rm +(P)}$ & $0.998_{-0.04}$ & $0.975_{-0.025}^{\rm +(P)}$ \\
$\alpha_{13}$ & $-0.8_{-0.2}^{+0.8}$ & $-0.6_{-0.3}^{+0.6}$ & $-0.3_{-0.4}^{+0.3}$ & $-1.0_{-0.3}^{+1.1}$ & $-0.1_{-0.7}^{+0.1}$ & $-0.5_{-0.4}^{+0.5}$ \\
$\dot{M}$ [$10^{18}$~g/s] & $0.56_{-0.04}^{+0.06}$ & $0.54_{-0.03}^{\rm +2.87}$ & $0.48_{-0.05}^{\rm +2.31}$ & $0.65_{-0.03}^{+0.67}$ & $0.81_{-0.18}^{+1.35}$ \\ 
$f_{\rm col}$ & $1.7^\star$ & $1.7^\star$ & $1.7^\star$ & $1.7^\star$ & $1.7^\star$ & $1.7^\star$ \\
\hline\hline
$\chi^2/\nu$ & 75.31/63 & 65.48/63 & 43.17/64 & 47.06/57 & 62.59/57 & 298.48/308 \\
& =1.195 & =1.039 & =0.675 & =0.826 & =1.098 & =0.969 \\
\hline\hline
\end{tabular}
}
\caption{\rm Best-fit values from the analysis of the \textsl{RXTE} observations of GRS~1915+105 with the model {\tt tbabs$\times$smedge$\times$gabs$\times$edge$\times$(comptt+nkbb)}. We show the results of the fits of the individual observations (R1, R2, R3, R4, and R5) and of the fit of all observations together (R1-R5, last column). The reported uncertainties correspond to 90\% confidence level for one relevant parameter ($\Delta\chi^2 = 2.71$). $^\star$ indicates that the parameter is frozen in the fit. (P) means that the 90\% confidence level limit is not in the parameter range allowed by the model (the maximum value of $a_*$ is 0.998, the minimum value of $W$ is 0.2~keV, and the minimum value of $kT_0$ is 0.01~keV). \label{t-rxte}}
\end{table*}

\subsection{\textsl{Suzaku} data}

GRS~1915+105 was observed by \textsl{Suzaku} for around 117~ks (Obs. ID 402071010) on 7 May 2007. Tab.~\ref{t-obs} shows the details of that observation. The source was in low/hard flux state. \citet{2009ApJ...706...60B} first reported the spectral analysis of this data and measured the BH spin $a_* > 0.98$ (assuming the Kerr metric).

\subsubsection{Data reduction}

Here we follow \citet{2020MNRAS.498.3565T} for data reduction. Out of the four XIS instruments, we use data only from XIS1 as two of the units were turned off during observation and the data from the fourth XIS unit is not suitable for spectral studies as it was running in timing mode during the observation. The unprocessed data of the aforementioned observation ID is downloaded from the HEASARC website and processed to filtered files using {\tt aepipeline} module,
distributed as part of the HEASOFT package, and  calibration database CALDB version 20151005. In order to minimize the effect of pile-up,
the source region is taken as an annulus with an inner radius of 78 arcsec and an outer radius of 208 arcsec. As background region, we select an annulus with the source at the center and the inner and outer radii of 208 and 278 arcsec, respectively. The redistribution matrix file is created using {\tt xisrmfgen} and the corresponding ancillary file is produced using {\tt xissimarfgen}. A net exposure of 29 ks is 
obtained after screening. Finally, the source spectrum is rebinned to a minimum of 25 counts per bin in order to apply $\chi^2$ statistics. 
We only consider data in the energy range 2.3-10.0 keV for spectral analysis in order to avoid calibration issues.

Data from the PIN detector are used for the HXD instrument. The raw data is processed using {\tt aepipeline}. The spectrum, ancillary file, and response
matrix file are obtained using the ftool {\tt hxdpinxbpi} and CALDB. The net exposure of 53 ks is obtained and the data is rebinned to 25 counts
per bin. For spectral analysis, data in the energy range 12.0-55.0 keV are used. During the spectral fits, the cross-calibration constant between
XIS1 and PIN data is kept free because of XIS1 being affected by pile-up.

GRS 1915+105 is believed to be a highly variable source but its flux and hardness did not vary much during this particular observation \citep[see][]{2020MNRAS.498.3565T}. 
The time-averaged spectrum is thus used without resolving the data in the time or flux domain.

\subsubsection{Data analysis}

The spectrum of this particular observation is simple and can be fit with coronal and reflection components, modified by galactic absorption \citep{2009ApJ...706...60B, 2020ApJ...899...80A}. In {\tt XSPEC}, the model reads as

\vspace{0.2cm}

{\tt tbabs$\times$(comptt+relxillCp\_nk)}

\vspace{0.2cm}

\noindent {\tt tbabs} describes the Galactic absorption \citep{2000ApJ...542..914W} and the column density $N_{\rm H}$ is left free in the fit. {\tt comptt} describes the Comptonization of seed photons in the hot plasma of the corona \citep{1994ApJ...434..570T}. The model includes three parameters which defines the Comptonized spectrum: the temperature of the seed photons $T_0$, the temperature of the plasma $T_{\rm e}$, and the plasma optical depth $\tau$. {\tt relxillCp\_nk} describes the reflection spectrum of the accretion disk in the Johannsen metric with possible non-vanishing deformation parameter $\alpha_{13}$ \citep{2017ApJ...842...76B, Abdikamalov:2019yrr}. We model the emissivity profile with a broken power-law, so we have three free parameters: the inner emissivity index $q_{\rm in}$, the outer emissivity index $q_{\rm out}$, and the breaking radius $R_{\rm br}$. Since we use {\tt comptt} to describe the continuum from the corona, the reflection fraction in {\tt relxillCp\_nk} is set to $-1$ and the model returns only the reflection spectrum. The electron temperature in {\tt relxillCp\_nk} is linked to $T_{\rm e}$ in {\tt comptt}.

\begin{figure}
\begin{center}
\includegraphics[width=8.5cm,trim={2.0cm 0.5cm 3.0cm 17.0cm},clip]{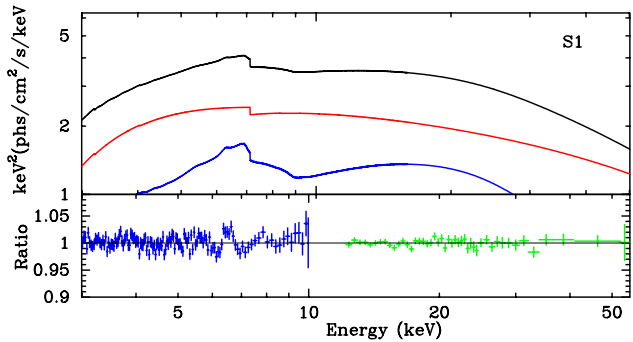}
\end{center}
\vspace{-0.4cm}
\caption{Best-fit model (upper quadrant) and data to the best-fit model ratio (lower quadrant) in the analysis of the \textsl{Suzaku} observation of GRS~1915+105. In the upper quadrant, we show the total spectrum (black curve), the {\tt relxillCp\_nk} component (red curve), and the {\tt comptt} component (blue curve). In the lower quadrant, blue crosses indicate the XIS data and green crosses are for the PIN data. \label{f-S-r}}
\end{figure}

The results of the fit are summarized in Tab.~\ref{t-suzaku}. Fig.~\ref{f-S-r} shows the best-fit model and the data to the best-fit model ratio.

\begin{table}
\centering
{\renewcommand{\arraystretch}{1.2}
\begin{tabular}{lc}
\hline\hline
Parameter & \hspace{1.0cm} Value \hspace{1.0cm} \\
\hline
{\tt tbabs} & \\
$N_{\rm H}$ [$10^{22}$~cm$^{-2}$] & $8.35^{+0.17}_{-0.18}$\\
\hline
{\tt comptt} & \\
$kT_0$ [keV] & $0.23_{-0.05}^{+0.08}$ \\
$kT_{\rm e}$ [keV] & $33_{-8}^{+8}$ \\
$\tau$ & $0.71_{-0.09}^{+0.22}$ \\
\hline
{\tt relxillCp\_nk} & \\
$q_{\rm in}$ & $8.9^{\rm + (P)}_{-1.4}$ \\
$q_{\rm out}$ & $0.00^{+0.21}$ \\
$R_{\rm br}$ [$r_{\rm g}$] & $7.8^{+3.4}_{-1.7}$ \\
$a_*$& $0.998_{-0.012}$ \\
$i$ [deg] &$73.3^{+3.0}_{-1.6}$\\
$\Gamma$ & $2.57^{+0.06}_{-0.05}$ \\
$\log\xi$ [erg~cm~s$^{-1}$] & $2.77^{+0.09}_{-0.06}$ \\
$A_{\rm Fe}$ & $0.82^{+0.13}_{-0.12}$\\
$\alpha_{13}$ & $0.00_{-0.10}^{+0.14}$\\
\hline
$C_{\rm HXD}$& $1.922^{+0.018}_{-0.024}$\\
\hline\hline
$\chi^2/\nu$ & 2281.51/2206 \\
& =1.03423 \\
\hline\hline
\end{tabular}
}
\caption{\rm Best-fit values from the analysis of the \textsl{Suzaku} observation of GRS~1915+105 with the model {\tt tbabs$\times$(comptt+relxillCp\_nk)}. The reported uncertainties correspond to 90\% confidence level for one relevant parameter ($\Delta\chi^2 = 2.71$). 
\label{t-suzaku}}
\end{table}

\begin{figure*}
\begin{center}
\includegraphics[width=8.5cm,trim={1.5cm 2.5cm 0.5cm 1.0cm},clip]{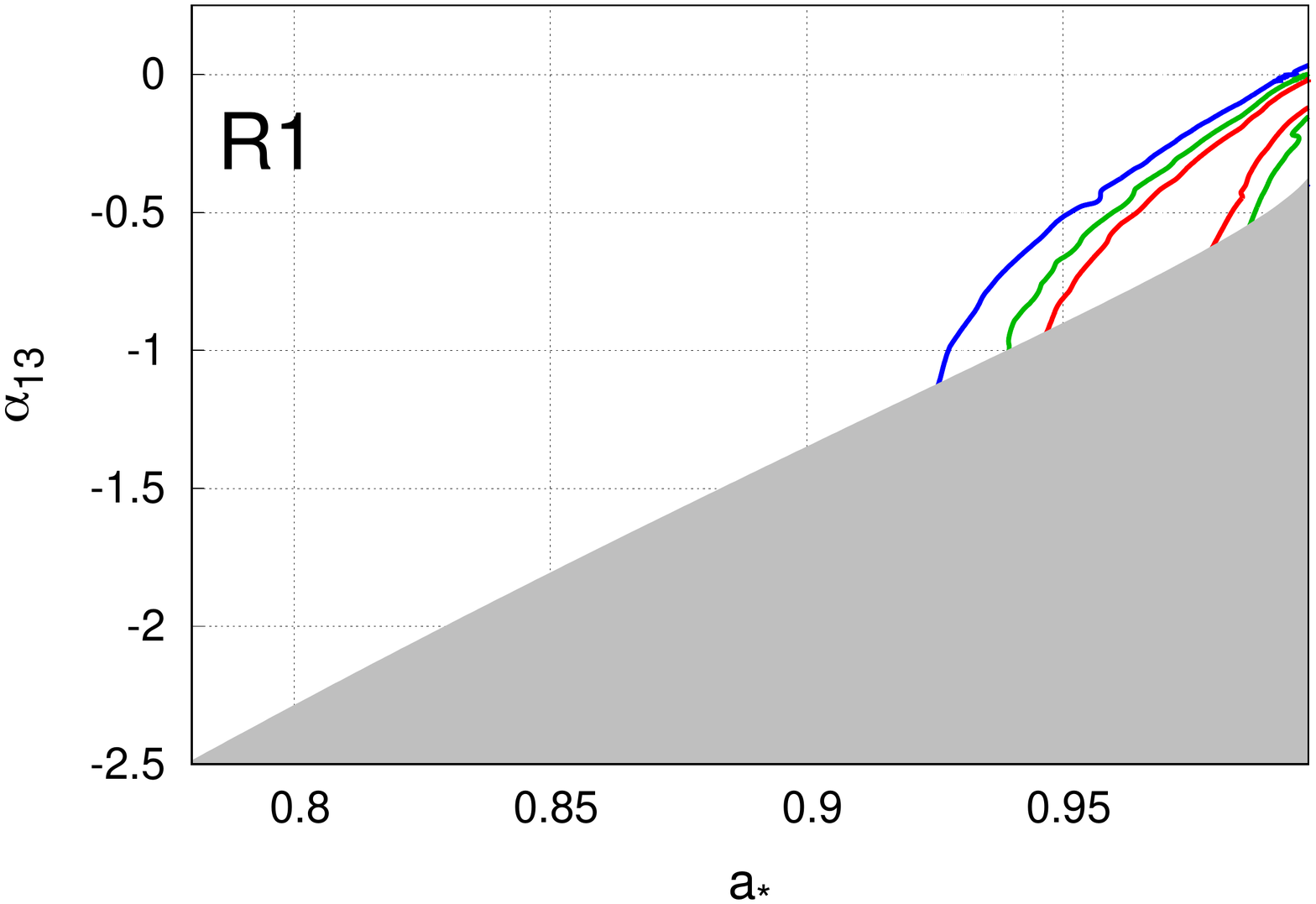}
\includegraphics[width=8.5cm,trim={1.5cm 2.5cm 0.5cm 1.0cm},clip]{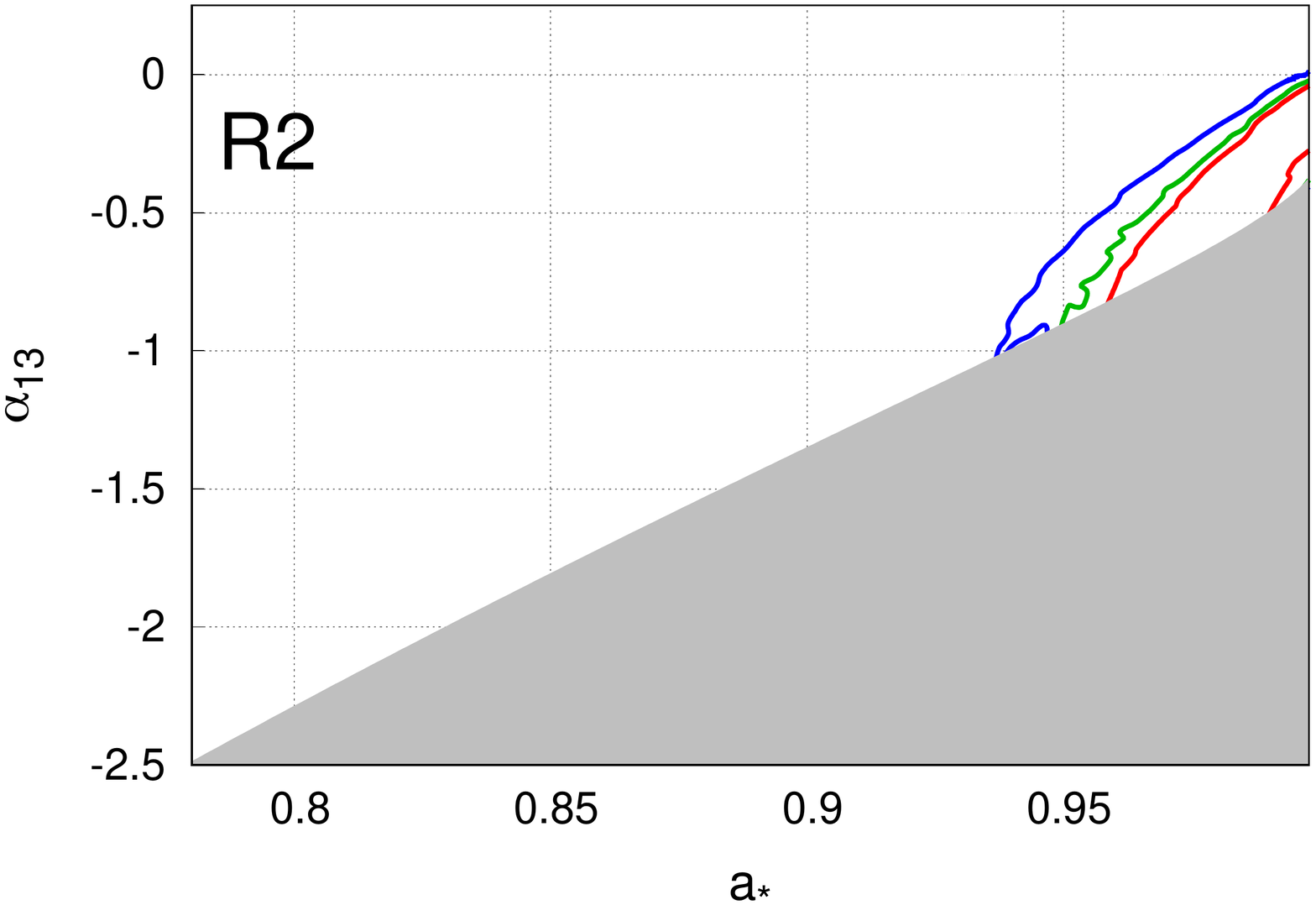} \\
\includegraphics[width=8.5cm,trim={1.5cm 2.5cm 0.5cm 1.0cm},clip]{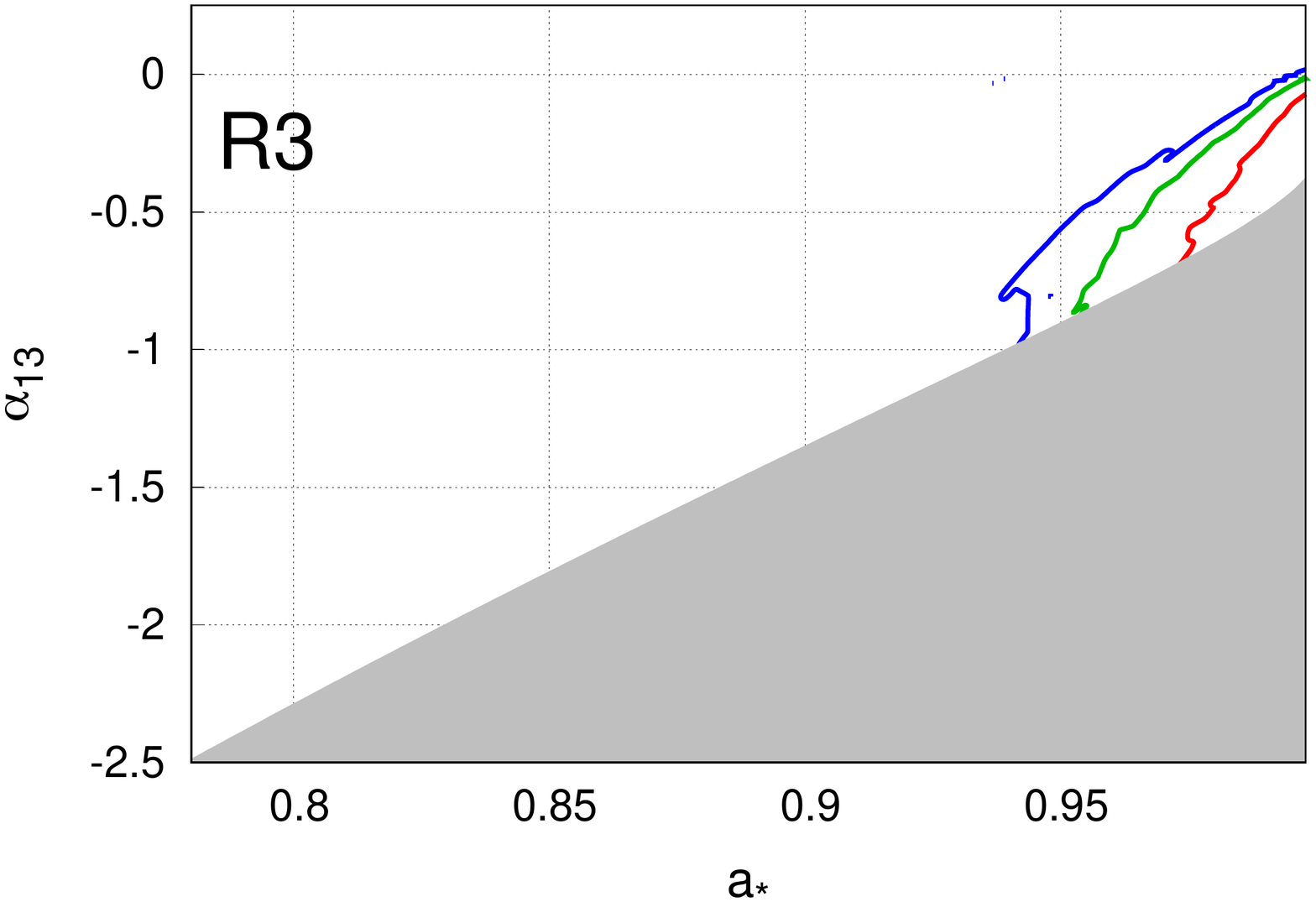}
\includegraphics[width=8.5cm,trim={1.5cm 2.5cm 0.5cm 1.0cm},clip]{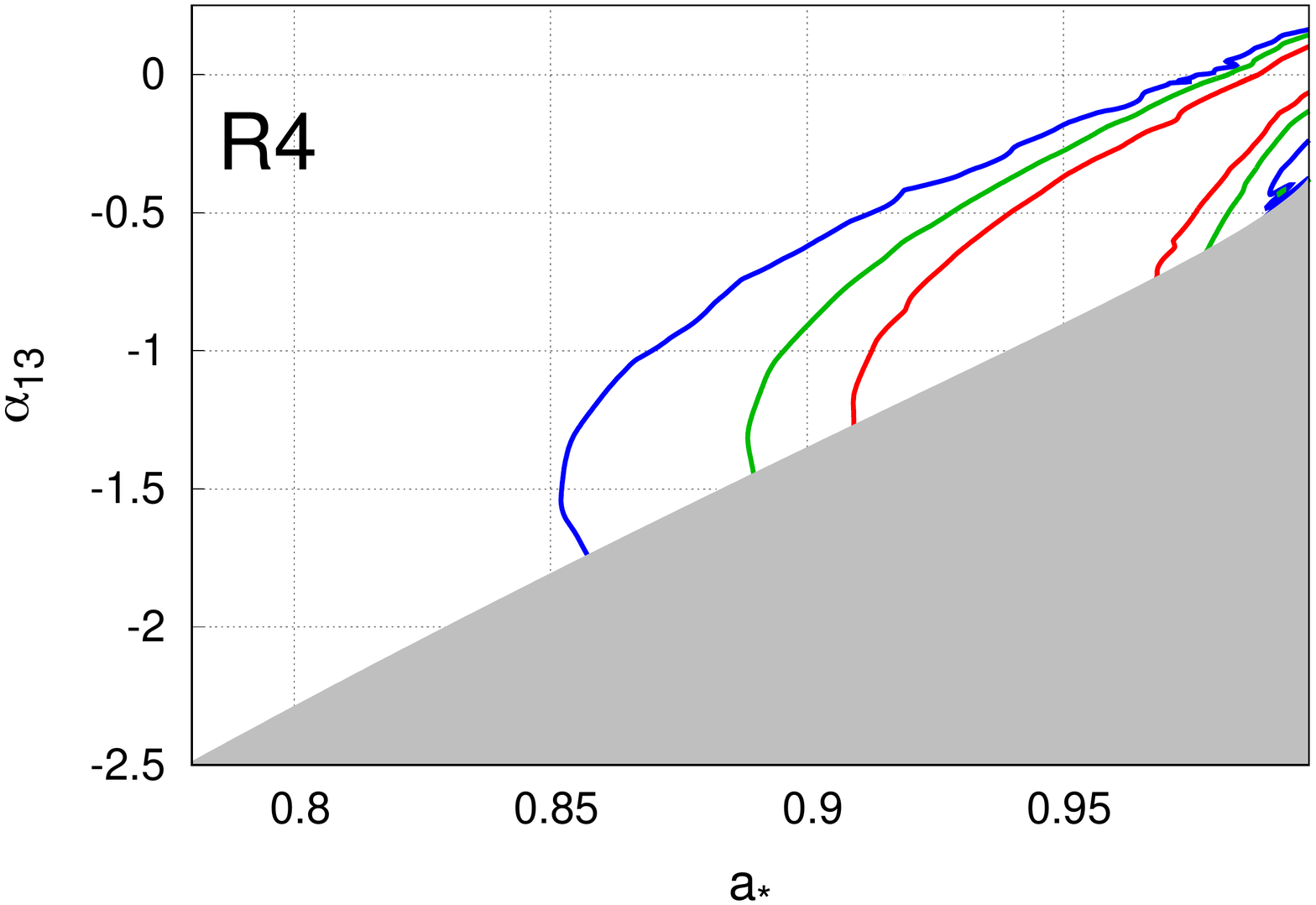} \\
\includegraphics[width=8.5cm,trim={1.5cm 2.5cm 0.5cm 1.0cm},clip]{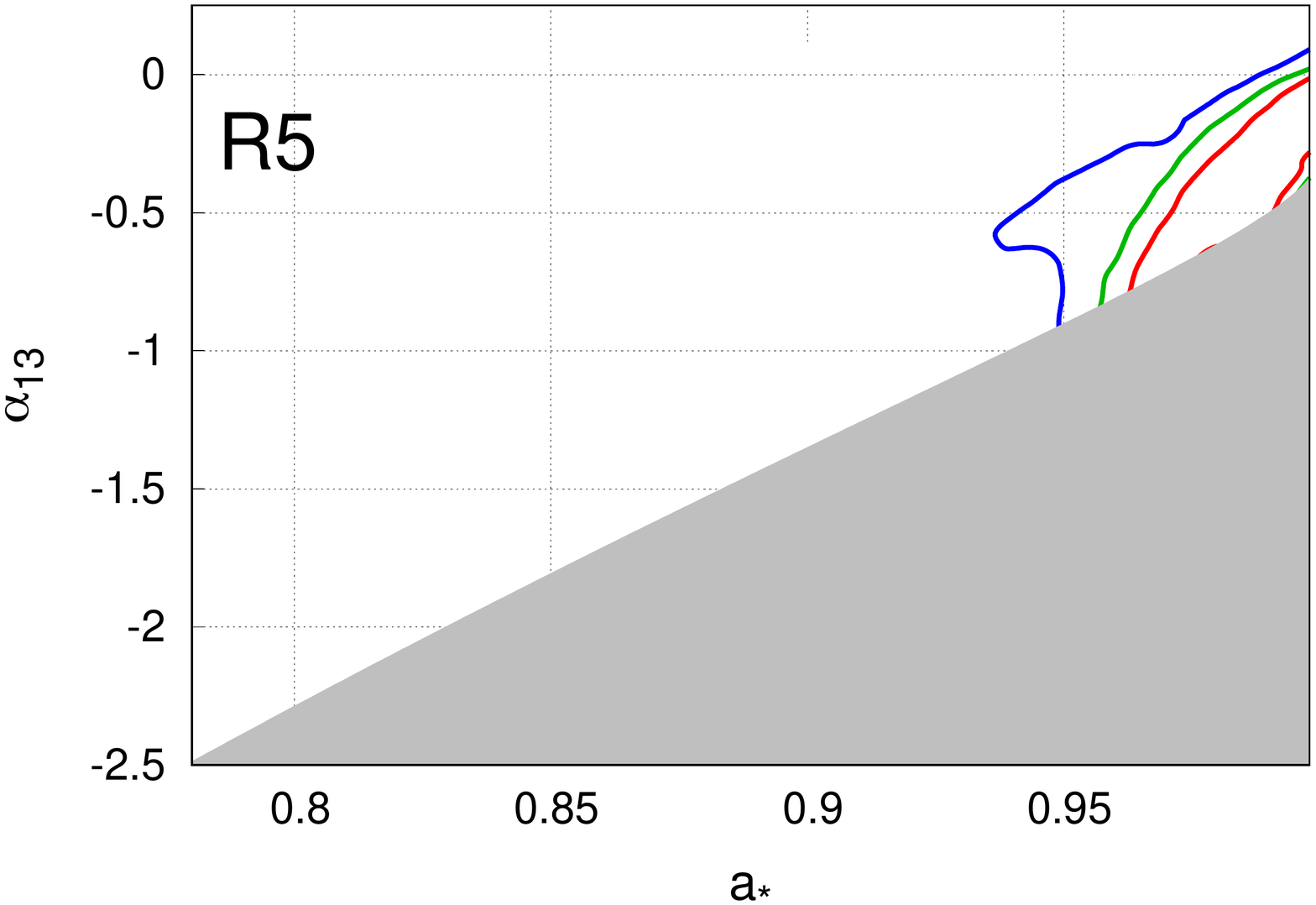}
\includegraphics[width=8.5cm,trim={1.5cm 2.5cm 0.5cm 1.0cm},clip]{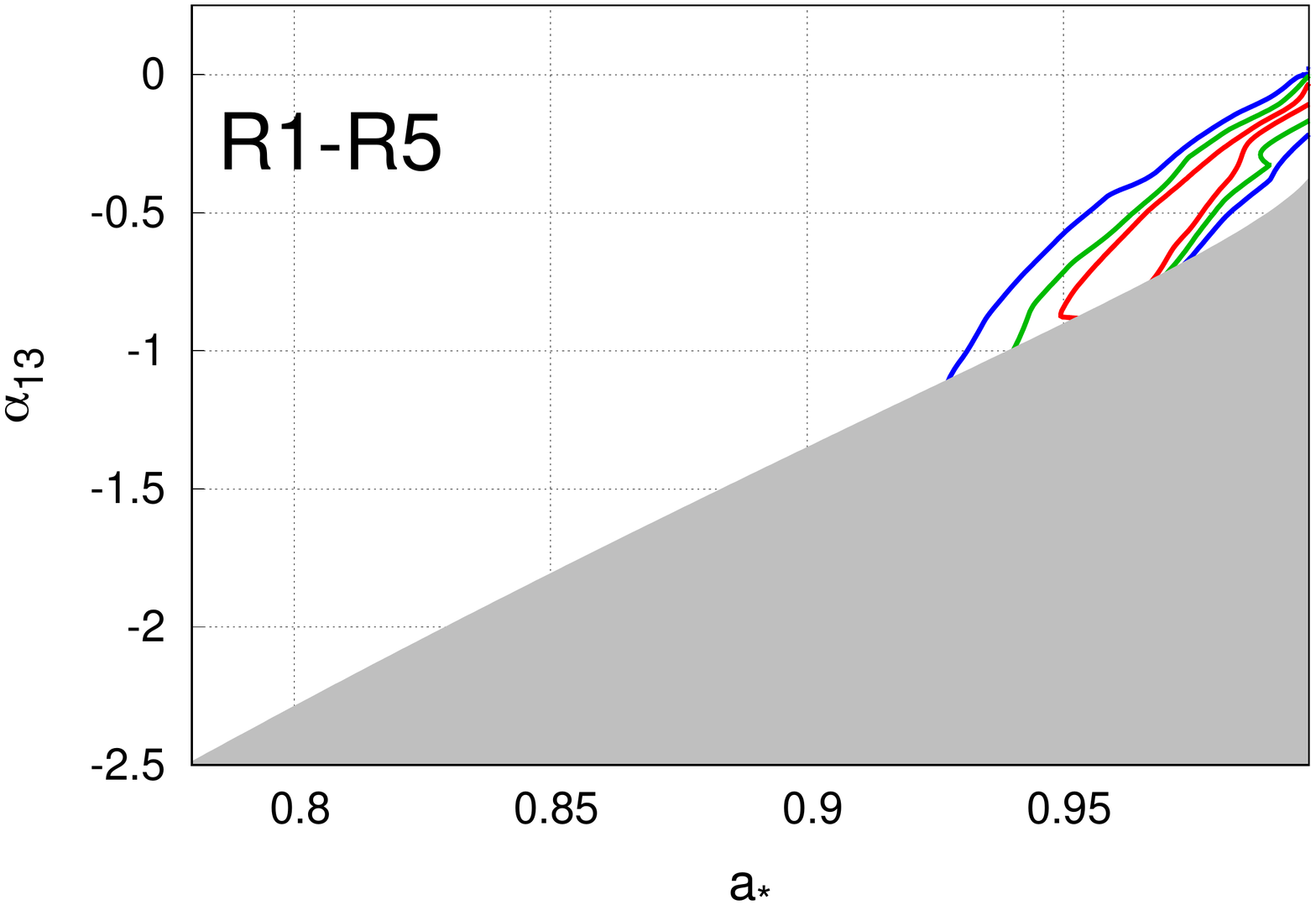}
\end{center}
\vspace{-0.2cm}
\caption{Constraints on the BH spin parameter $a_*$ and the deformation parameter $\alpha_{13}$ from the individual fits of the five \textsl{RXTE} observations of GRS~1915+105 (R1, R2, R3, R4, and R5) and from the fit of all observations together (R1-R5, bottom-right panel). The red, green, and blue curves represent, respectively, the 68\%, 90\%, and 99\% confidence level limits for two relevant parameters ($\Delta\chi^2 = 2.30$, 4.61, and 9.21, respectively). The gray region is ignored in our analysis because it violates the constraint in Eq.~(\ref{eq-constraints}) and the spacetime presents pathological properties there. \label{f-rxte}}
\end{figure*}

\begin{figure}
\begin{center}
\includegraphics[width=8.5cm,trim={1.5cm 2.5cm 0.5cm 1.0cm},clip]{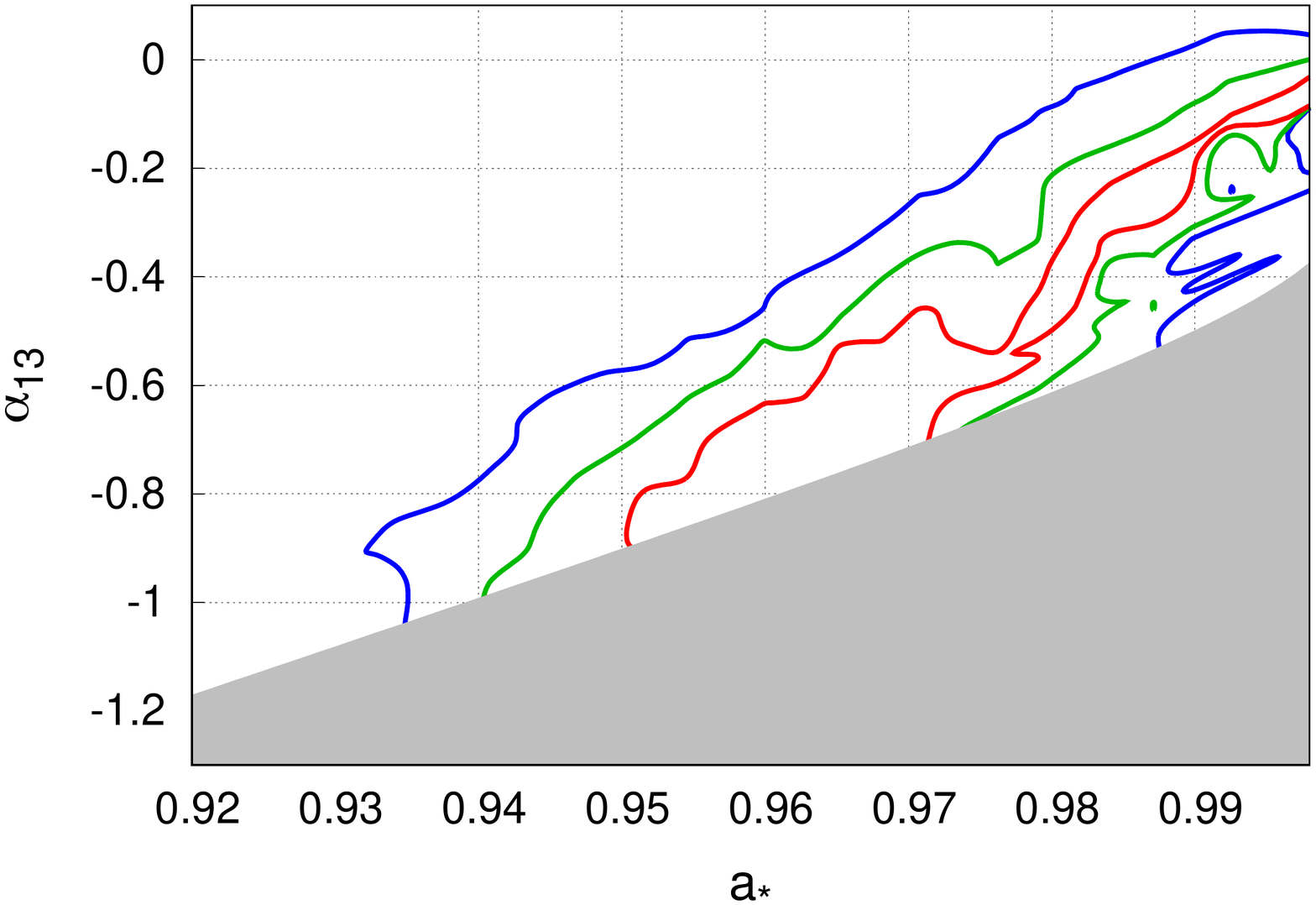}
\end{center}
\vspace{-0.2cm}
\caption{{Constraints on the BH spin parameter $a_*$ and the deformation parameter $\alpha_{13}$ from the fit of the five \textsl{RXTE} observations together and taking the uncertainties on the BH mass, distance, and inclination angle of the disk into account by using the $\chi^2$ in Eq.~(\ref{eq-chi2222}). The red, green, and blue curves represent, respectively, the 68\%, 90\%, and 99\% confidence level limits for two relevant parameters ($\Delta\chi^2 = 2.30$, 4.61, and 9.21, respectively). The gray region is ignored in our analysis because it violates the constraint in Eq.~(\ref{eq-constraints}) and the spacetime presents pathological properties there.} \label{f-rxte-mdi}}
\vspace{0.7cm}
\end{figure}


\section{Discussion and conclusions}
\label{sec:discussion}

In the present paper, we have analyzed five disk-dominated \textsl{RXTE} data and one reflection-dominated \textsl{Suzaku} spectrum of the BH in GRS~1915+105 employing the models {\tt nkbb} and {\tt relxill\_nk}. Our goal is to test the Kerr hypothesis by combining the study of the thermal component and the reflection features. We note that similar studies were reported in \citet{2021ApJ...907...31T} and in \citet{2021arXiv210603086Z}. \citet{2021ApJ...907...31T} analyzed a \textsl{NuSTAR} observation of GX~339--4 in which spectrum simultaneously shows a strong thermal component and prominent reflection features. However, the BH mass and the distance of the source are not known for GX~339--4, so they were inferred in the fit {while the continuum-fitting method requires to know the values of these two parameters in order to estimate the BH spin (and the possible deformation parameter). \citet{2021arXiv210603086Z} analyzed three \textsl{NuSTAR} observations of the BH in GRS~1716--249 in the hard-intermediate state where, again, we simultaneously see a strong thermal spectrum and prominent reflection features. For GRS~1716--249, the distance is known, while for the BH mass we only have a lower limit and its value was thus inferred by the fit. The case of GRS~1915+105 studied in the present work is thus somewhat different. The BH mass and distance are both known \citep{2014ApJ...796....2R}. Moreover, we have five \textsl{RXTE} observations in which we see a very strong thermal component and we do not see any reflection feature, and one \textsl{Suzaku} spectrum with strong reflection features and a negligible thermal component.

Let us start with the five \textsl{RXTE} observations analyzed with {\tt nkbb}. First, we analyzed every observation separately (R1, R2, R3, R4, and R5) and then we fit the data of all observation together (fit R1-R5). The best-fit values of these six fits are reported in Tab.~\ref{t-rxte}. We note that the line energy in  {\tt edge}, {\tt smedge}, and {\tt gabs} are consistent among the five individual fits. We always recover a high value of the BH spin parameter, while the estimate of the deformation parameter includes the Kerr metric $\alpha_{13} = 0$ but allows for negative values of $\alpha_{13}$. Fig.~\ref{f-rxte} shows the constraints on $a_*$ and $\alpha_{13}$ of the six fits. The combined analysis of the five observations provide the following constraint on $\alpha_{13}$ (90\% confidence level)
\begin{eqnarray}
\alpha_{13} = -0.5^{+0.5}_{-0.4} \, .
\end{eqnarray}
This constraint is somewhat weaker than the typical constraints obtained from the analysis of the reflection features with {\tt relxill\_nk}; see, e.g., Tab.~5 in \citet{2021arXiv210603086Z}. However, it is stronger than the constraint on $\alpha_{13}$ inferred by the study of the thermal spectrum of LMC~X-1 with {\tt nkbb} in \citet{Tripathi:2020qco}. While in both studies (here and for LMC~X-1) we clearly see a strong correlation between the estimate of $a_*$ and $\alpha_{13}$, here we recover a higher value of the BH spin, which helps to constrain the deformation parameter $\alpha_{13}$ because the parameter space is smaller (the gray region in Fig.~\ref{f-rxte} is ignored in our analysis because the spacetimes have pathological properties there; see the appendix for more details).

We note that the fits in Tab.~\ref{t-rxte} and the constraints in Fig.~\ref{f-rxte} are obtained by freezing the values of the BH mass $M$, viewing angle $\iota$, and source distance $D$ as shown in Tab.~\ref{t-rxte}. However, these three quantities are known with a finite uncertainty. {In we plug a lower/higher value for the BH mass $M$ (without changing the values of $\iota$ and $D$), the valley of the minimum of $\chi^2$ moves to the left/right in the $a_*$ vs. $\alpha_{13}$ plane. Similarly, if we employ a lower/higher value of the inclination angle $\iota$ (without changing the values of $M$ and $D$), the valley of the minimum of $\chi^2$ moves to the right/left in the $a_*$ vs. $\alpha_{13}$ plane. Last, if we use a lower/higher value of the source distance $D$ (without changing the values of $M$ and $\iota$), the valley of the minimum of $\chi^2$ moves to the right/left in the $a_*$ vs. $\alpha_{13}$ plane. In order to take the uncertainties of $M$, $\iota$, and $D$ into account, we can leave these three parameters free in {\tt nkbb} and we calculate the $\chi^2$ as
\be\label{eq-chi2222}
\chi^2_{\rm tot} = \chi^2_{\rm RXTE} + \frac{\left(M_* - M\right)^2}{\sigma_M^2} + \frac{\left(\iota_* - \iota\right)^2}{\sigma_\iota^2} + \frac{\left(D_* - D\right)^2}{\sigma_D^2} \, , \nonumber\\
\ee 
where $\chi^2_{\rm RXTE}$ is the $\chi^2$ contribution from the \textsl{RXTE} data; $M$, $\iota$, and $D$ are free parameters in the model; $M_* = 12.4~M_\odot$, $\sigma_M = 2~M_\odot$, $\iota = 65^\circ$, $\sigma_\iota = 10^\circ$, $D_* = 8.6$~kpc, and $\sigma_D = 2.0$~kpc, following the measurements of $M$, $D$, and $\iota$ in \citet{2014ApJ...796....2R} and \citet{1999MNRAS.304..865F}. The last three terms on the right hand side of Eq.~(\ref{eq-chi2222}) are intended to incorporate, in a very simple way, the priors on those parameters. With this procedure, we find the constraints on $a_*$ and $\alpha_{13}$ reported in Fig.~\ref{f-rxte-mdi}.}

The best-fit values from the \textsl{Suzaku} fit are reported in Tab.~\ref{t-suzaku}. This observation was already analyzed with an older version of {\tt relxill\_nk} in \citet{2019ApJ...884..147Z} and therefore we do not further discuss this fit here. We note that the 90\% confidence level measurement of the Johannsen deformation parameter $\alpha_{13}$ is
\be
\alpha_{13} = 0.00_{-0.10}^{+0.14} \, .
\ee
This is quite a stringent constraint, especially if compared with the constraints from the five disk-dominated spectra inferred with {\tt nkbb}. While every constraint depends on the quality of the data and the spectrum of the source during a particular observation, it is surely true that the analysis of the reflection features is a more powerful method to probe the strong gravity region around a BH than the continuum-fitting method. The thermal spectrum has indeed a very simple shape, it is just a multi-temperature blackbody spectrum, which can be fit well by a number of different combinations of the model parameters. The constraint on the BH spin and the deformation parameter from the analysis of this \textsl{Suzaku} spectrum are shown in Fig.~\ref{f-suzaku}.

\begin{figure*}
\begin{center}
\includegraphics[width=8.5cm,trim={1.5cm 2.5cm 0.5cm 1.0cm},clip]{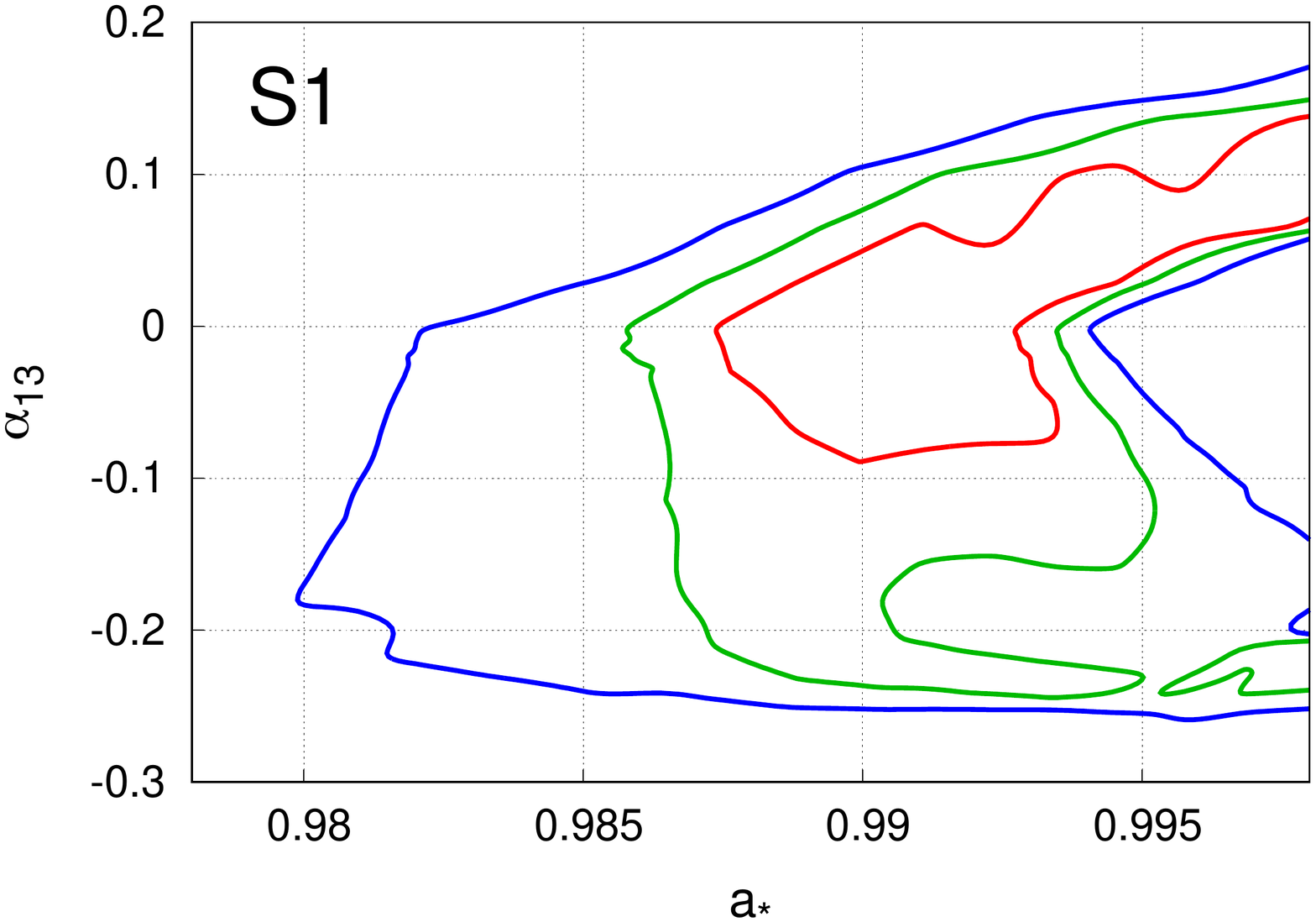}
\end{center}
\vspace{0.0cm}
\caption{Constraints on the BH spin parameter $a_*$ and the deformation parameter $\alpha_{13}$ from the \textsl{Suzaku} observation of GRS~1915+105. The red, green, and blue curves represent, respectively, the 68\%, 90\%, and 99\% confidence level limits for two relevant parameters ($\Delta\chi^2 = 2.30$, 4.61, and 9.21, respectively). \label{f-suzaku}}
\end{figure*}

\begin{figure*}
\begin{center}
\includegraphics[width=8.5cm,trim={1.5cm 2.5cm 0.5cm 1.0cm},clip]{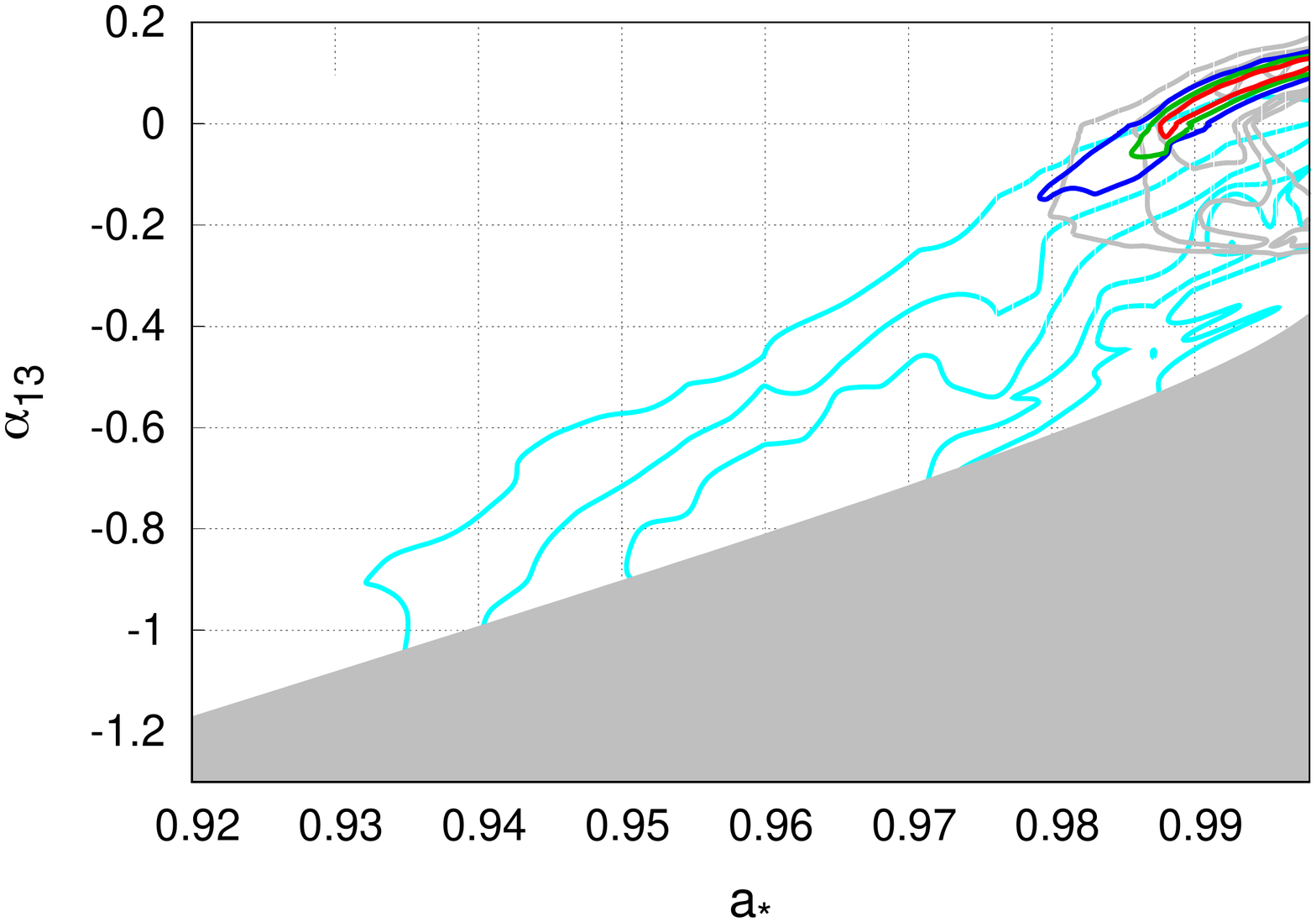}
\includegraphics[width=8.5cm,trim={1.5cm 2.5cm 0.5cm 1.0cm},clip]{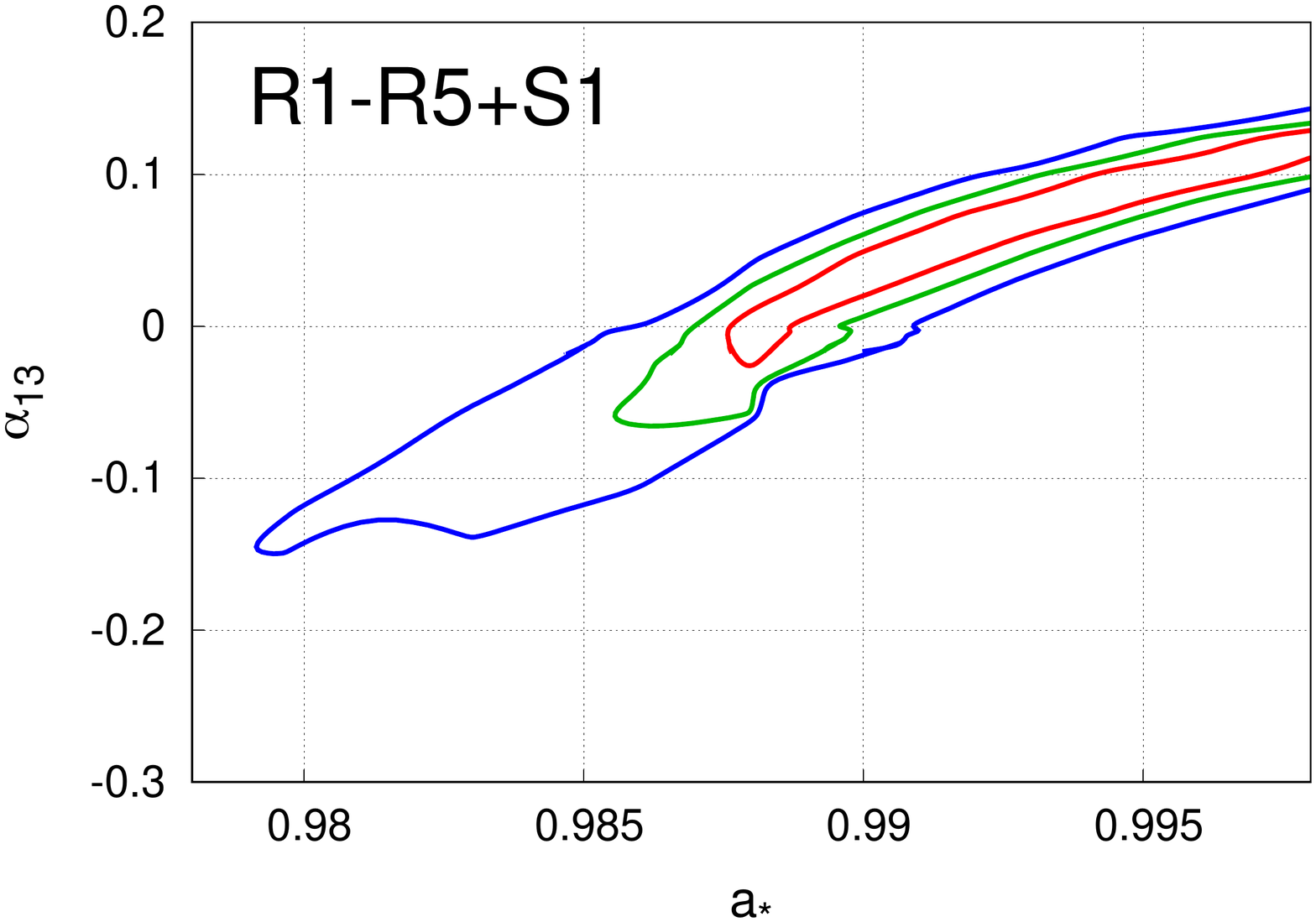}
\end{center}
\vspace{0.0cm}
\caption{{The left panel shows the constraints on the BH spin parameter $a_*$ and the deformation parameter $\alpha_{13}$ from the \textsl{RXTE} data (cyan curves and previously shown in Fig.~\ref{f-rxte-mdi}), the \textsl{Suzaku} data (gray curves and previously shown in Fig.~\ref{f-suzaku}), and the combined analysis of the \textsl{RXTE} and \textsl{Suzaku} observations (where the red, green, and blue curves represent, respectively, the 68\%, 90\%, and 99\% confidence level limits for two relevant parameters). The right panel shows only the combined analysis of the \textsl{RXTE} and \textsl{Suzaku} observations.} \label{f-suzaku-rxte}}
\end{figure*}

{Last, we want to combine the analysis of the \textsl{RXTE} and \textsl{Suzaku} data together\footnote{{Since the \textsl{RXTE} and \textsl{Suzaku} observations are not simultaneous, we do not have cross-calibration constants between the \textsl{RXTE} and \textsl{Suzaku} instruments. However, this does not introduce any calibration uncertainty. The normalization of the flux is not important in the analysis of the \textsl{Suzaku} data because all details are in the shape of the spectrum (and, indeed, we can analyze the data without any knowledge of the BH mass and distance). For the analysis of the thermal spectrum in the \textsl{RXTE} data, the normalization of the flux is important, but here the uncertainty is dominated by the uncertainties on the BH mass and distance, while the calibration uncertainty is subdominant.}}, taking the uncertainties on $M$ and $D$ into account (the value of the viewing angle $\iota$ can be determined by the fit of the reflection features in the \textsl{Suzaku} spectrum and the parameter $\iota$ in {\tt nkbb} will be linked to that in {\tt relxill\_nk}). We proceed as before and we calculate the $\chi^2$ as
\be\label{eq-chi2345}
\chi^2_{\rm tot} = \chi^2_{\rm RXTE} + \chi^2_{\rm Suzaku} + \frac{\left(M_* - M\right)^2}{\sigma_M^2} + \frac{\left(D_* - D\right)^2}{\sigma_D^2} \, , \nonumber\\
\ee 
where $\chi^2_{\rm RXTE}$ and $\chi^2_{\rm Suzaku}$ are, respectively, the $\chi^2$ contribution from the \textsl{RXTE} and \textsl{Suzaku} data, $M$ and $D$ are free parameters in the model, $M_* = 12.4~M_\odot$ $\sigma_M = 2~M_\odot$, $D_* = 8.6$~kpc, and $\sigma_D = 2.0$~kpc following the measurements of $M$ and $D$ in \citet{2014ApJ...796....2R}. As in the analysis of the \textsl{RXTE} data, the last two terms on the right hand side of Eq.~(\ref{eq-chi2345}) are intended to incorporate the uncertainties on $M$ and $D$ in quite a simple way. The fit leads to the constraints on $a_*$ and $\alpha_{13}$ shown in Fig.~\ref{f-suzaku-rxte}.} 
We recover Kerr and the constraints on $a_*$ and $\alpha_{13}$ are only slightly better than the constraints inferred from the sole analysis of the \textsl{Suzaku} spectrum. The estimate of $\alpha_{13}$ is (90\% confidence level)
\be
\alpha_{13} = 0.12_{-0.14}^{+0.02} \, .
\ee
As we have already stressed, the analysis of the reflection features can indeed provide much stronger constraints than the continuum-fitting method. However, the combination of the two techniques can improve the constraints and, more importantly, it represents an important check for the systematics, which is often difficult to take into account.

\begin{figure*}
\begin{center}
\includegraphics[width=16.0cm,trim={1.0cm 2.0cm 3.0cm 17.5cm},clip]{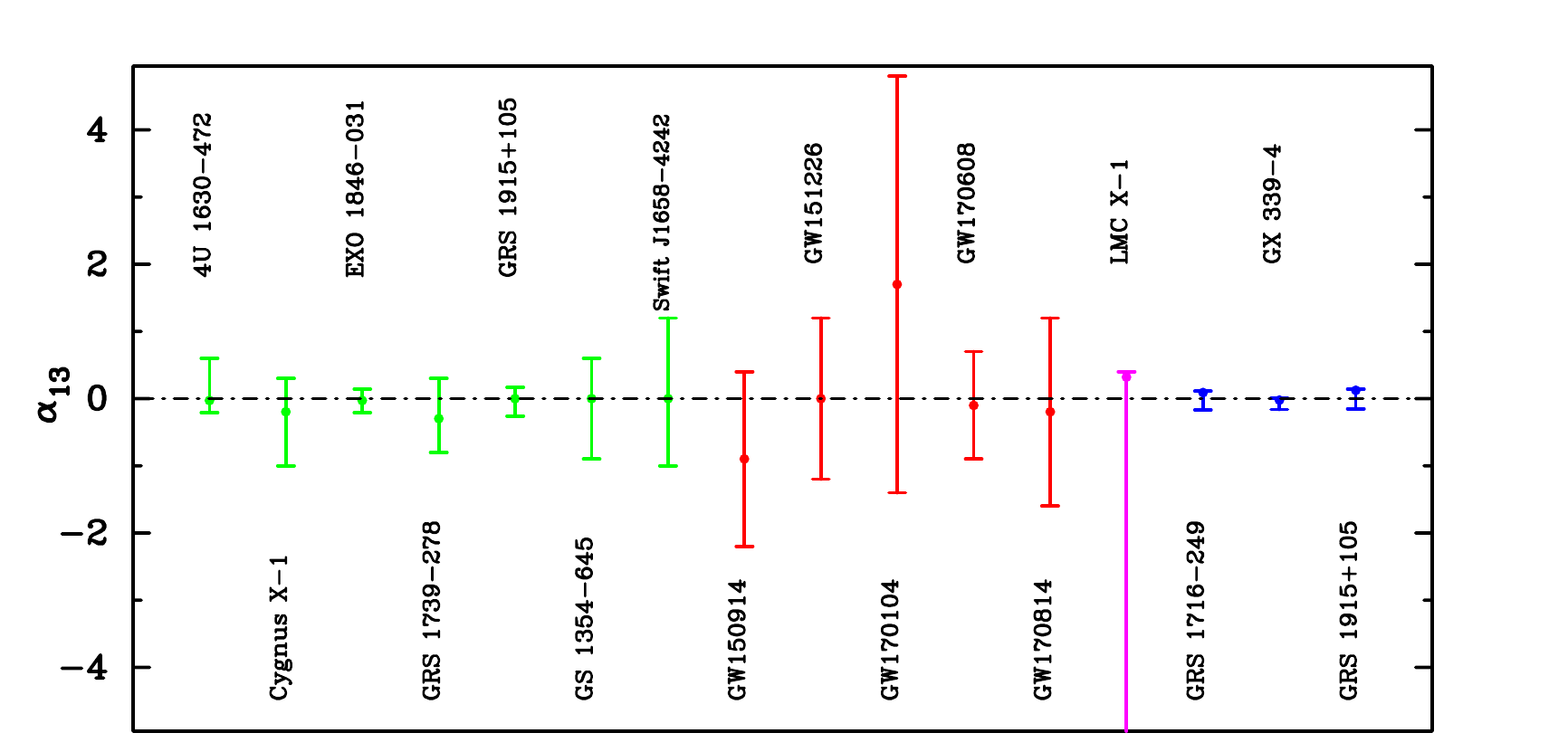}
\end{center}
\vspace{-0.2cm}
\caption{Summary of the available constraints on the Johannsen deformation parameter $\alpha_{13}$ from stellar-mass BHs. All error bars show the 3-$\sigma$ measurement. The constraints inferred from the analysis of reflection features are in green, those from gravitational wave data are in red, that from the continuum-fitting method in magenta, and the constraints obtained from the combination of the analysis of reflection features and the disk's thermal spectrum are in blue. See the text for more details. \label{f-summary}}
\vspace{0.5cm}
\end{figure*}

Fig.~\ref{f-summary} shows all the currently available constraints on the deformation parameter $\alpha_{13}$ from stellar-mass BHs with different techniques\footnote{The references of the original measurements can be found in \citet{2021arXiv210603086Z}.}: the constraints are from the analysis of reflection features with {\tt relxill\_nk} (green), from gravitational wave data \citep{2020CQGra..37m5008C} (red), with the continuum-fitting method with {\tt nkbb} (magenta), and from the analysis of reflection features and disk's thermal spectrum by using {\tt relxill\_nk} and {\tt nkbb}. The sole analysis of the disk's thermal component (the case of LMC~X-1 in Fig.~\ref{f-summary}) cannot constrain $\alpha_{13}$ well because of the strong degeneracy between $a_*$ and $\alpha_{13}$. However, the combination of the analysis of reflection features and disk's thermal spectrum can improve the constraining power of X-ray reflection spectroscopy and from Fig.~\ref{f-summary} we can see that the constraints in blue are the most stringent ones. We could imagine to further improve these constraints if we could test the spacetime metric around accreting BHs with the measurement of the frequencies of QPOs. As of now, this is not yet possible, because the exact mechanism responsible for QPOs is unknown. However, most QPO models involve the spacetime metric and it is thus likely that in the future, once the origin of these QPOs will be understood, we will also be able to combine the QPO data to further strengthen the constraints on the deformation parameters.

\vspace{0.5cm}

{\bf Acknowledgments --}
This work was supported by the Innovation Program of the Shanghai Municipal Education Commission, Grant No.~2019-01-07-00-07-E00035, the National Natural Science Foundation of China (NSFC), Grant No.~11973019, and Fudan University, Grant No.~JIH1512604.
D.A. is supported through the Teach@T{\"u}bingen Fellowship.
A.T., C.B., H.L., and V.G. are members of the International Team~458 at the International Space Science Institute (ISSI), Bern, Switzerland, and acknowledge support from ISSI during the meetings in Bern.


\appendix

\section{Johannsen metric}\label{a:johannsen}

For convenience of the reader, we report here the expression of the Johannsen metric. More details can be found in the original paper \citet{Johannsen:2015pca}. In Boyer-Lindquist-like coordinates, the line element is
\be\label{eq-jm}
ds^2 &=&-\frac{\tilde{\Sigma}\left(\Delta-a^2A_2^2\sin^2\theta\right)}{B^2}dt^2
+\frac{\tilde{\Sigma}}{\Delta A_5}dr^2+\tilde{\Sigma} d\theta^2 
-\frac{2a\left[\left(r^2+a^2\right)A_1A_2-\Delta\right]\tilde{\Sigma}\sin^2\theta}{B^2}dtd\phi \nonumber\\
&&+\frac{\left[\left(r^2+a^2\right)^2A_1^2-a^2\Delta\sin^2\theta\right]\tilde{\Sigma}\sin^2\theta}{B^2}d\phi^2
\ee
where $M$ is the BH mass, $a = J/M$, $J$ is the BH spin angular momentum, $\tilde{\Sigma} = \Sigma + f$, and
\be
\Sigma = r^2 + a^2 \cos^2\theta \, , \qquad
\Delta = r^2 - 2 M r + a^2 \, , \qquad
B = \left(r^2+a^2\right)A_1-a^2A_2\sin^2\theta \, .
\ee
The functions $f$, $A_1$, $A_2$, and $A_5$ are defined as
\be
f = \sum^\infty_{n=3} \epsilon_n \frac{M^n}{r^{n-2}} \, , \quad
A_1 = 1 + \sum^\infty_{n=3} \alpha_{1n} \left(\frac{M}{r}\right)^n \, , \quad
A_2 = 1 + \sum^\infty_{n=2} \alpha_{2n}\left(\frac{M}{r}\right)^n \, , \quad
A_5 = 1 + \sum^\infty_{n=2} \alpha_{5n}\left(\frac{M}{r}\right)^n \, ,
\ee
where $\{ \epsilon_n \}$, $\{ \alpha_{1n} \}$, $\{ \alpha_{2n} \}$, and $\{ \alpha_{5n} \}$ are four infinite sets of deformation parameters without constraints from the Newtonian limit and Solar System experiments. In the present study, we only consider the deformation parameter $\alpha_{13}$, which has the strongest impact on the reflection spectrum, and we assume that all other deformation parameters vanish.

In order to avoid pathological properties in the spacetime, we have to impose some constraints on the values of $a_*$ and $\alpha_{13}$. For the BH spin, we require $| a_* | \le 1$. This is the same constraint as in the Kerr spacetime: for $| a_* | > 1$, there is no event horizon and the metric describes the spacetime of a naked singularity. For the deformation parameter $\alpha_{13}$, we require
\be
\label{eq-constraints}
\alpha_{13} > - \frac{1}{2} \left( 1 + \sqrt{1 - a^2_*} \right)^4 \, .
\ee


\end{document}